\documentclass[10pt,twocolumn,twoside]{IEEEtran}
\usepackage[mathcal]{euscript}
\usepackage{amssymb}
\usepackage{amsfonts}
\usepackage{amsmath}
\usepackage{cite}
\usepackage[dvips]{graphicx}
\usepackage{color}
\usepackage{comment}
\usepackage{url}

\begin{document}

\newtheorem{definition}{\it Definition}
\newtheorem{theorem}{\bf Theorem}
\newtheorem{lemma}{\it Lemma}
\newtheorem{corollary}{\it Corollary}
\newtheorem{remark}{\it Remark}
\newtheorem{example}{\it Example}
\newtheorem{case}{\bf Case Study}
\newtheorem{assumption}{\it Assumption}
\newtheorem{property}{\it Property}
\newtheorem{proposition}{\it Proposition}

\newcommand{\hP}[1]{{\boldsymbol h}_{{#1}{\bullet}}}
\newcommand{\hS}[1]{{\boldsymbol h}_{{\bullet}{#1}}}
\newcommand{\ba}{\boldsymbol{a}}
\newcommand{\baq}{\overline{q}}
\newcommand{\bA}{\boldsymbol{A}}
\newcommand{\bcA}{\boldsymbol{\cal A}}
\newcommand{\bb}{\boldsymbol{b}}
\newcommand{\bB}{\boldsymbol{B}}
\newcommand{\bc}{\boldsymbol{c}}
\newcommand{\bC}{\boldsymbol{C}}
\newcommand{\bcC}{\boldsymbol{\cal C}}
\newcommand{\bcE}{\boldsymbol{\cal E}}
\newcommand{\bcO}{\boldsymbol{\cal O}}
\newcommand{\bd}{\boldsymbol{d}}
\newcommand{\be}{\boldsymbol{e}}
\newcommand{\bh}{\boldsymbol{h}}
\newcommand{\bH}{\boldsymbol{H}}
\newcommand{\bl}{\boldsymbol{l}}
\newcommand{\bL}{\boldsymbol{L}}
\newcommand{\bm}{\boldsymbol{m}}
\newcommand{\bn}{\boldsymbol{n}}
\newcommand{\bo}{\boldsymbol{o}}
\newcommand{\bO}{\boldsymbol{O}}
\newcommand{\bp}{\boldsymbol{p}}
\newcommand{\bq}{\boldsymbol{q}}
\newcommand{\br}{\boldsymbol{r}}
\newcommand{\bR}{\boldsymbol{R}}
\newcommand{\bs}{\boldsymbol{s}}
\newcommand{\bS}{\boldsymbol{S}}
\newcommand{\bT}{\boldsymbol{T}}
\newcommand{\bu}{\boldsymbol{u}}
\newcommand{\bw}{\boldsymbol{w}}
\newcommand{\bcY}{\boldsymbol{\cal Y}}
\newcommand{\bx}{\boldsymbol{x}}
\newcommand{\by}{\boldsymbol{y}}

\newcommand{\balpha}{\boldsymbol{\alpha}}
\newcommand{\bbeta}{\boldsymbol{\beta}}
\newcommand{\bOmega}{\boldsymbol{\Omega}}
\newcommand{\bTheta}{\boldsymbol{\Theta}}
\newcommand{\bphi}{\boldsymbol{\phi}}
\newcommand{\btheta}{\boldsymbol{\theta}}
\newcommand{\btau}{\boldsymbol{\tau}}
\newcommand{\bvarpi}{\boldsymbol{\varpi}}
\newcommand{\bpi}{\boldsymbol{\pi}}
\newcommand{\bpsi}{\boldsymbol{\psi}}
\newcommand{\bxi}{\boldsymbol{\xi}}
\newcommand{\blambda}{\boldsymbol{\lambda}}

\newcommand{\cA}{{\cal A}}
\newcommand{\cB}{{\cal B}}
\newcommand{\cC}{{\cal C}}
\newcommand{\cD}{{\cal D}}
\newcommand{\cE}{{\cal E}}
\newcommand{\cF}{{\cal F}}
\newcommand{\cG}{{\cal G}}
\newcommand{\cH}{{\cal H}}
\newcommand{\bcH}{\boldsymbol {\cal H}}
\newcommand{\cK}{{\cal K}}
\newcommand{\cM}{{\cal M}}
\newcommand{\cO}{{\cal O}}
\newcommand{\cP}{{\cal P}}
\newcommand{\cQ}{{\cal Q}}
\newcommand{\cR}{{\cal R}}
\newcommand{\cS}{{\cal S}}
\newcommand{\dcS}{\ddot{{\cal S}}}
\newcommand{\ds}{\ddot{{s}}}
\newcommand{\cT}{{\cal T}}
\newcommand{\cU}{{\cal U}}
\newcommand{\cV}{{\cal V}}
\newcommand{\cW}{{\cal W}}
\newcommand{\cY}{{\cal Y}}
\newcommand{\wt}[1]{\widetilde{#1}}

\newcommand{\mA}{\mathbb{A}}
\newcommand{\mE}{\mathbb{E}}
\newcommand{\mG}{\mathbb{G}}
\newcommand{\mR}{\mathbb{R}}
\newcommand{\mS}{\mathbb{S}}
\newcommand{\mU}{\mathbb{U}}
\newcommand{\mV}{\mathbb{V}}
\newcommand{\mW}{\mathbb{W}}

\newcommand{\uq}{\underline{q}}
\newcommand{\ubq}{\underline{\boldsymbol q}}

\newcommand{\red}[1]{\textcolor[rgb]{1,0,0}{#1}}
\newcommand{\gre}[1]{\textcolor[rgb]{0,1,0}{#1}}
\newcommand{\blu}[1]{\textcolor[rgb]{0,0,0}{#1}}

\title{Dynamic Network Slicing for Scalable Fog Computing Systems with Energy Harvesting} 

\author{Yong~Xiao, \IEEEmembership{Senior Member, IEEE} and Marwan Krunz, \IEEEmembership{Fellow, IEEE}

\thanks{

Y. Xiao is with the School of Electronic Information
and Communications at the Huazhong University of Science and Technology, Wuhan, China (e-mail: yongxiao@hust.edu.cn).

M. Krunz is with the Department of Electrical and Computer Engineering at the University of Arizona, Tucson, AZ (e-mail: krunz@email.arizona.edu). M. Krunz is also with the University Technology Sydney. }
}

\maketitle
\vspace{-0.8in}
\begin{abstract}
This paper studies fog computing systems,  
in which cloud data centers can be supplemented by a large number of fog nodes deployed in a wide geographical area. Each node relies on harvested energy from the surrounding environment to provide computational services to local users.
We propose the concept of {\em dynamic network slicing} 
in which a regional orchestrator coordinates workload distribution among local fog nodes, 
providing partitions/slices of energy and computational resources to support a specific type of service with certain quality-of-service (QoS) guarantees. 
The resources allocated to each slice can be dynamically adjusted according to service demands and energy availability. 
A stochastic overlapping coalition-formation game is developed to investigate distributed cooperation and joint network slicing between fog nodes under randomly fluctuating energy harvesting and workload arrival processes. We observe that the overall processing capacity of the fog computing network can be improved by allowing fog nodes to 
maintain a belief function about the unknown state and  the private information of other nodes. 
An algorithm based on a belief-state partially observable Markov decision process (B-POMDP) is proposed to achieve the optimal resource slicing structure among all fog nodes. 
We describe how to implement our proposed dynamic network slicing within the 3GPP network sharing architecture, and  
%
evaluate the performance of our proposed framework using the real BS location data 
of a real cellular system with over 200 BSs deployed in the city of Dublin. 
Our numerical results show that our framework can significantly improve the workload processing capability of fog computing networks. In particular, even when each fog node can  coordinate only with its closest neighbor, the total amount of workload processed by fog nodes can be almost doubled under certain scenarios.
\end{abstract}



\vspace{-0.1in}
\section{Introduction}
\label{Section_Introduction}
With the widespread proliferation of intelligent systems, Internet-of-Things (IoT) devices, and smart infrastructures, computation-intensive mobile applications that require low delay and fast processing time are becoming quite popular. Next-generation mobile networks (e.g., 5G and beyond) are expected to serve over 50 billion mobile devices, most of which are smart devices requiring as low as 1 millisecond latency and very little energy consumption\blu{\cite{ITU2014TactileInternet, Huawei5GVision,Andrews5G}}. Major IT service providers, such as Google, Yahoo, Amazon, etc., are heavily investing in large-scale data centers to meet the demand for future data services. 
%
However, these data centers are expensive and often built in remote areas to save costs. 
This makes it difficult to provide the quality-of-service (QoS) requirements of end users, especially for users located at the edge of a coverage area.
%
%
To provide low-latency services to end users, a new framework referred to as {\em fog computing} has emerged\cite{Vaquero2014FogComp}, in which a large number of wired/wireless, closely located, and often decentralised devices, commonly referred to as {\em fog nodes}, can communicate and potentially cooperate with each other to perform certain computational tasks. Fog computing complements existing cloud services by distributing computation, communication, and control tasks closer to end users. Fog nodes 
include a variety of devices between end users and data centers, such as routers, smart gateways, access points (APs), base stations (BSs), and set-top boxes. According to Next-Generation Mobile Network (NGMN) Alliance\cite{NGMN5GWhitePaper}, fog computing will be an important component of 5G systems, providing support for computation-intensive applications that require low latency, high reliability, and secure services. Examples of these applications include intelligent transportation, smart infrastructure, e-healthcare, and augmented/virtual reality (AR/VR).
%
The success of fog computing heavily relies on the ubiquity and intelligence of low-cost fog nodes to reduce the latency and relieve network congestion\cite{Dastjerdi2016FogComp, Yi2015Fog, Yannuzzi2014Fog}.

Over the last decade, there has been a significant interest in climate change and energy sustainability for information and communication technologies\blu{\cite{GoogleGreenCloud, AppleGreenCloud, MicrosoftGreenCloud, FacebookGreenCloud}}. The telecommunication network infrastructure is already one of the leading sources of global carbon dioxide emissions\cite{Chamola2016SolarBS}. In addition, the unavailability of a reliable energy supply from electricity grids in some areas is forcing mobile network operators (MNOs) to use sources like diesel generators for power, which not only increase operating costs but also contribute to pollution. 
Energy harvesting is a technology that allows electronic devices to be powered by the energy converted from the environment, such as sunlight, wind power, and tides. It has recently attracted significant interest due to its potential to provide a sustainable energy source for electronic devices with zero carbon emission\cite{Ulukus2015EHreview, XY2015DET, Lu2015EHSurvey, XY2015ICCBayeRL, Ge2015EnergyEff}. 
\blu{Major IT providers including Apple, Facebook, and Google have already upgraded all their cloud computing servers to be fully supported by renewable energy\cite{FacebookGreenCloud,MicrosoftGreenCloud,GoogleGreenCloud}. }
Allowing fog nodes to utilize the energy harvested from the Nature can provide 
ubiquitous computational resources anywhere at any time. For example, fog nodes deployed inside an edge network can rely on renewable energy sources to support low-latency, real-time computation for applications such  as environmental control, traffic monitoring and congestion avoidance, automated real-time vehicle guidance systems, and AR/VR assisted manufacturing. 

Incorporating energy harvesting into the design of the fog computing infrastructure is still relatively unexplored. In contrast to data centers that can be supported by massive photovoltaic solar panels or wind turbines, fog nodes are often limited in size and location. 
In addition, it is generally difficult to have a global resource manager that  coordinates resource distribution among fog nodes in a centralized fashion. Developing a simple and effective method for fog nodes to optimize their energy and computational resources, enabling autonomous resource management according to the time-varying energy availability and user demands, is still an open problem. 
%

Enabled by software-defined networking (SDN) and network function virtualization (NFV) technologies,  the concept of {\em network slicing}
has recently been introduced by 3GPP to further improve the flexibility and scalability of fog computing for 5G systems\cite{3GPP2016NetworkShare,3GPP2016NetworkShare2,Vaezi2017VirtualizationCloud,NGMN2016NetworkSlicing}.
Network slicing allows a fog node to support multiple types of service (use cases) by partitioning its resources, such as spectrum, infrastructure, network functionality, and computing power among these types. Resource partitions, commonly referred to as {\em slices}, can be tailored and orchestrated according to different QoS requirements of different service types (e.g., real-time audio, image/video processing with various levels of delay tolerance, etc.)\cite{Richart2016NetSlicing}. 
Multiple SDN-based network slicing architectures have been proposed by 3GPP\cite{Samdanis2016NetworkSlicingBroker}, 
NGMN Alliance\cite{NGMN5GWhitePaper,NGMN2016NetworkSlicing}, and Open Networking Foundation (ONF) \cite{ONF5GSlicing, ONFSDNArchitecture}. However, these architectures are all based on a centralized control plane and cannot be directly applied to large-scale network systems.

In this paper, we introduce a new {\em dynamic network slicing} architecture for large-scale energy-harvesting fog computing networks. 
This architecture embodies a new network entity, the {\em regional SDN-based orchestrator}, that coordinates the workload processed by multiple closely located fog nodes and creates slices 
of energy and computational resources for various types of service requested by end users. 
To minimize the coordination cost, the workload of each user is first sent to the closest fog node. Fog nodes will then make autonomous decisions on how much energy resource is spent on activating computational resources and how to partition the activated computational resources according to time-varying energy availability, user demands, and QoS requirements. If a fog node decides that it needs 
help from its neighboring fog nodes to process a part of its received workload, or if it has surplus resource to 
help other fog nodes in proximity, it will coordinate with these fog nodes though the regional SDN-based orchestrator.
Our main objective is to develop a simple distributed network slicing policy that can maximize the utilization efficiency of available resources and balance the workloads among fog nodes over a wide geographical area.
The distributed and autonomous decision making process at each fog node makes game theory a suitable tool to analyze the interactions among fog nodes. In this paper, we develop a stochastic overlapping coalition-formation game-based framework, called {\em dynamic network slicing game}, to analyze such interactions. In contrast to the traditional partition-based coalition formation game, in our game, players are allowed to interact with each other across multiple coalitions, which has the potential to further improve the resource utilization efficiency and increase the outcome for players. Unfortunately, finding a stable coalitional structure in this game is known to be notoriously difficult. Because each player can allocate a fraction of its resources to each coalition, there can be infinitely many possible coalitions among players. It has already been proved that an overlapping coalition game may not always have a stable coalitional structure. Even it does, there is no general method that can converge to such a structure.  
We propose a distributed algorithm based on a belief-state partially observable Markov decision process (B-POMDP) for each fog node to sequentially learn from its past experience and update its belief function about the state and offloading capabilities of other nodes. We prove that our proposed algorithm can achieve the optimal resource slicing policy without requiring back-and-forth communication between fog nodes. 
Finally,  
we evaluate the performance of our proposed framework by simulations, using actual BS topological deployment of 
a large-scale cellular network in the city of Dublin.
Results show that our proposed framework can significantly improve the workload offloading capability of fog nodes. In particular, even when each fog node can only cooperate with its closest neighbor, the total amount of workload offloaded by the fog nodes can almost be doubled especially for densely deployed fog nodes in urban areas.

\vspace{-0.1in}
\section{Related Work}
\label{Section_RelatedWork}
A key challenge for fog computing is to provide QoS-guaranteed computational services to end users 
while optimizing the utilization of local resources owned by fog nodes\cite{Sarkar2016Fog, Bonomi2014Fog, Datta2015Fog}. 
In \cite{Do2015Fog}, the joint optimization of allocated resources while minimizing the carbon footprint 
was studied for video streaming services over fog nodes. 
A service-oriented resource estimation and management model was proposed in \cite{Aazam2015Fog} for fog computing systems.  In \cite{XY2018TactielInternet}, a distributed optimization algorithm has been proposed for fog computing-supported Tactile Internet applications requiring ultra-low latency services.

\blu{SDN and NFV have been considered as key enablers for fog computing\cite{Harshit2016SDFog, Khan2018VANETFog}. In particular, popular SDN protocols such as ONF's OpenFlow have already been extended into fog computing networks\cite{Hakiri2017Fog}. To Further improve the scalability and flexibility of OpenFlow when extending to wireless systems, the authors in \cite{Tootoonchian2010SDNHyperflow} introduced a hybrid SDN control plane that combines the Optimized Link State Routing Protocol (OLSR), a popular IP routing protocol for mobile ad hoc network, with the OpenFlow to perform path searching and selection as well as network monitoring. Recently, a new framework, referred to as the hierarchical SDN, has been introduced to reduce the implementation complexity of SDN by organizing the network components such as controllers and switchers in a layered structure\cite{Fang2015HierarchicalSDN}. In particular, the authors in \cite{Hakiri2017Fog} proposed a hierarchical framework, called hyperflow, in which groups of switches have been assigned to controllers to keep the decision making within individual controllers. The authors in \cite{Yeganeh2012KandooSDN} developed a new control plane consisting of two layers of controllers: bottom-layer and top-layer controllers. The former runs only locally controlled  applications without inter-connections nor the knowledge  of  the  network-wide state. And the latter corresponds to a logically centralized controller that maintains the network-wide state. In \cite{Liu2015HierarchicalSDN}, the authors investigated the optimization of the hierarchical organization for a set of given network scales. It shows that using a 4-layer SDN is sufficient for most practical network scales.}

Recently, game theory has been shown to be a promising tool to analyze the performance and optimize fog computing networks. Specifically, in \cite{ZhangHQ2016FogComp}, a hierarchical game-based model was applied to analyze the interactions between cloud data centers (CDCs) and fog nodes. An optimal pricing mechanism was proposed for CDCs to control resource utilization at fog nodes. 

\blu{To the best of our knowledge, our paper is the first work to study the distributed workload offloading and resource allocation problem for energy harvesting fog computing networks with fog node cooperation. }

\vspace{-0.1in}
\section{SDN-based Dynamic Network Slicing}
\label{Section_SDNArchitecture}

\subsection{
Fog Computing}
A generic fog computing architecture consists of the following elements\cite{Chiang2017FogBook,Zhang2017FogCompIoT,Zhang2017FogCompTCC,Tong2016}:

\begin{itemize}
\item[1)] {\em Cloud Computing Service Provider (CSP)} -- The CSP owns and manages large-scale CDCs that 
    can provide abundant hardware and software resources with low processing delay. 
    CDCs are often built in low-cost remote areas and therefore services processed at the CDC are expected to experience high transmission latency.  
\item[2)] {\em Fog Computing Service Provider (FSP)} -- The FSP controls a large number of low-cost fog nodes (e.g., mini-servers), deployed in a wide geographical area. Typically, fog nodes do not have high-performance processing units. 
    However, they are much cheaper to deploy and require much less energy to operate. In this paper, we focus on energy-harvesting fog computing networks in which computational resources that can be activated 
    are time-varying and solely depend on the harvested energy. 

\item[3)] {\em Networking Service Provider (NSP)} -- The NSP deploys a large wired or wireless network infrastructure that connects users to fog nodes and/or remote CDCs. 

\item[4)] \blu{{\em Tenants} -- Tenants can correspond to virtual network operators (VNOs) that lack network infrastructure or with limited capacity and/or coverage,
and have to lease resources from other service/infrastructure providers. They can also be over-the-top (OTT) service/content providers, such as Netflix, Spotify, Skype, etc., operating on top of the hardware/software infrastructures deployed by CSP, FSP, and/or NSP. In this paper, we assume each tenant always requests networking and/or computational resources (e.g., slices) from one or more providers 
to serve the needs for the users.} 

\item[5)] {\em Users} -- Users are mobile devices or user applications that consume the services offered by tenants. Users can locate in a wide geographical area and can request different types of services with different QoS requirements.
\end{itemize}

\blu{Note that the above elements may not always be physically separated from one another. 
For example, a cellular network operator (an NSP) with insufficient computational resources can rent computational resources (e.g., server/CPU times) from a FSP to support computational intensive service (e.g., AR/VR-based services, online gaming, etc.) requested by its subscribers\cite{Hadzic2017EdgeComp}. In this case, the NSP will also be considered as a tenant of the computational infrastructure of the FSP. Similarly, CSP/FSP can also rent networking infrastructure of an NSP to reduce the service response-time of its users. In this case, the CSP/FSP will be the tenant of the networking infrastructure of the NSP. The interaction between tenants and service providers (e.g., NSP/CSP/FSP) is closely related to the ownership as well as availability of the shared resources. In this paper, we consider a decentralized architecture and assume the tenants, CSP, NSP, and FSP are associated with different providers/operators. }  

In this paper, we mainly focus on the resource slicing/partitioning of the computational resources of the FSP and assume each tenant can always obtain sufficient networking resources from the NSP to deliver the requests of users to the intended fog servers.

\vspace{-0.1in}
\subsection{Existing Network Slicing Architectures}
A comprehensive SDN architecture was introduced by ONF in \cite{ONFSDNArchitecture}. In this architecture, an intermediate control plane 
is used to deliver tailored services to users in the application plane by configuring and abstracting the physical resources. 
The proposed SDN architecture can naturally support network slicing\cite{ONF5GSlicing}. In particular, the SDN controller is supposed to collect all the information needed to communicate with each user and create a complete abstract set of resources (as resource groups) to support control logic that constitutes a slice, including the complete collection of related service attributes of users. 
Although ONF's SDN architecture provides a comprehensive view of the control plane functionalities that enable network slicing, its centralized nature cannot support scalable deployment.

In \cite{Sciancalepore2017NetSlice, Caballero2017NetSlicingGame, Bega2017NetworkSlicing, Samdanis2016NetworkSlicingBroker}, the authors introduced the concept of 5G network slicing broker in 3GPP service working group SA 1. The proposed concept is based on the 3GPP's network sharing management architecture\cite{3GPP2016NetworkShare, 3GPP2016NetworkShare2}. 
In this concept, each tenant can acquire slices from service providers to run network functions.
In contrast to the ONF's network slicing architecture in which the slicing is created by 
exchanging resource usage information between tenants and service providers,
the 5G network slicing broker allows the service providers to directly create network slices according to the requirement of tenants.  
It can therefore support on-demand resource allocation and admission control. However, the network slicing broker only supports slicing of networking resources that are centrally controlled by a master operator-network manager (MO-NM)\footnote{According to 3GPP's network sharing management architecture\cite{3GPP2016NetworkShare,3GPP2016NetworkShare2}, to ensure optimized and secure resource allocation, a single master operator must be assigned as the only entity to centrally monitor and control shared network resources. }.

\blu{There are many other architectures that can also be extended to support network slicing. However, in terms of the entities (either each tenant directly requests the service slices from the FSP, or tenants and FSP must negotiate and jointly decide the slicing/partitioning of the resource) that need to be involved when making network slicing decisions, these architectures can be considered as the special cases of ONF and 3GPP's architectures. }

\begin{figure}
\centering
\includegraphics[width=1.8 in]{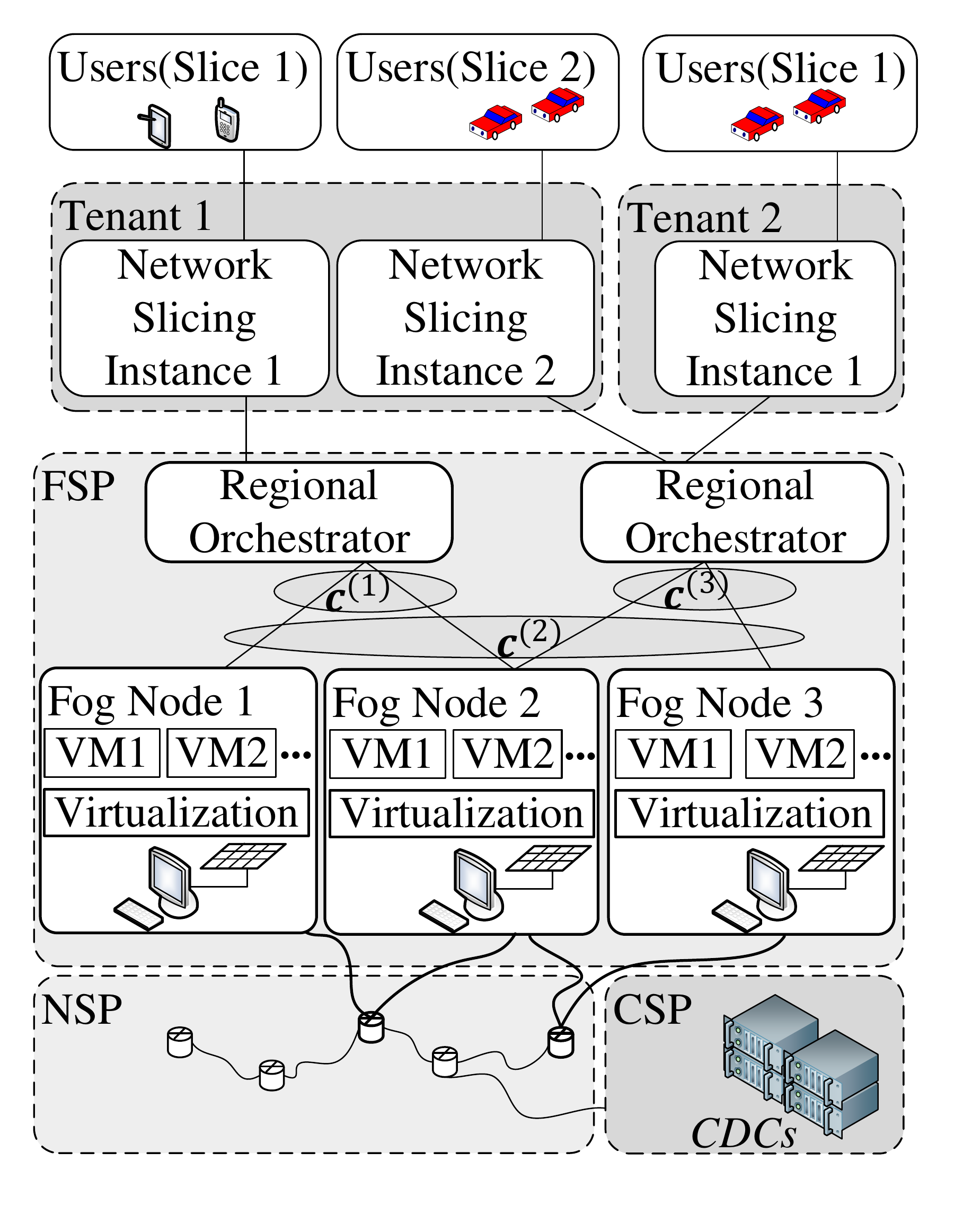}
\vspace{-0.1 in}
\caption{Dynamic network slicing architecture. }
\label{Figure_RegOrchSDNSlicing}
\vspace{-0.2in}
\end{figure}

\subsection{Regional SDN-based Orchestrator}
We propose the dynamic network slicing architecture that supports large-scale fog computing network on a new entity, the regional  orchestrator. In our architecture, each fog node $i$ coordinates with a subset of its neighboring fog nodes $\cC_i$ via a regional  orchestrator to create network slices for a  common set of services requested by the local users. More specifically, each tenant sends the resource request 
together with the location information of each user.
The workload request of each user is first assigned to the closest fog node. Each fog node will then partition its own resources according to the total received requests. If a fog node receives requests that exceed its available resources, it will coordinate with the regional  orchestrator to outsource a part of its load to one or more neighboring fog nodes. Similarly, if a fog node has surplus resources, 
it will report this surplus to the regional orchestrator, who 
will then coordinate with other fog nodes 
to forward the appropriate amounts of their workload to the nodes with surplus resource. 
Our proposed architecture is illustrated in Figure \ref{Figure_RegOrchSDNSlicing}.

\vspace{-0.2in}
\section{Problem Formulation}
\label{Section_SystemModel}

We consider a fog computing network \blu{consisting of a set of $N$ fog nodes, labeled as ${\cal F} = \{1, 2,\ldots, N\}$. 
Each fog node $i$ serves a set ${\cal B}_i$ of tenants (e.g., VNO or service/content providers) located in its coverage area. 
Different types of service can have different QoS requirements, in here measured by the service response-time. Each tenant can support at least one type of service.} 
Examples of service types that require ultra-low response-time ($<$10 ms) and high computational resources include AR/VR as well as traffic guidance and planning services for high-speed (self-driving) vehicles\cite{ATT2017EdgeNetworks, IETF2017ARVRResponsetime}. Other services that are more latency-tolerant  (e.g., $\approx$100 ms) include speech recognition and language translation. Let $\cV$ be the set of $K$ types of service supported by the FSP. Each tenant can request a subset of the service types in $\cV$. We use superscript $k$ to denote the parameters related to the $k$th service type. Let $\theta^{(k)}$ be the maximum tolerable service response-time for type $k$ service. 
We assume the tenants can always obtain sufficient networking resources from the NSP to deliver their workload to fog nodes and receive feedback results. Our methods can be directly extended to support slicing of both communication and computational resources, e.g., if network resources obtained by each tenant can only support a portion of the users' workload, then the tenant will only request slices of computational resources to process this particular portion of workload.

\subsection{Resource Constraints}
We consider an energy-harvesting fog computing network in which the workload processing service for each fog node is powered by 
the harvested energy. We assume time-varying (slotted) energy harvesting and workload arrival processes as in \cite{XY2015DET, Aprem2013EnergHarv}. Each fog node can harvest different amounts of energy and receive different amounts of workload from its users during different time slots. However, the workload arrival rate as well as the computational resource activated by each fog node are assumed to be constant within each time slot. We follow a commonly adopted setting\cite{Keller2014QDelay,Deng2016FogCompu} and assume that the workload arrival process for each service type $k$ in each time slot $t$ at each tenant $b\in {\cal B}_i$ follows a Poisson process ${\cal P} \left( \tilde \lambda^{(k)}_{i[b],t} \right)$ where $\lambda^{(k)}_{i[b],t}$ is average number of requests received in time slot $t$. Accordingly, 
the aggregated workload arrival rate for each service type $k$ at each fog node $i$ also follows a Poisson distribution ${\cal P} \left( \lambda^{(k)}_{i,t} \right)$, where $\lambda^{(k)}_{i,t} = \sum_{b\in {\cal B}_i} \tilde \lambda^{(k)}_{i[b],t}$. 
We focus on the computational resource partition/slicing and energy scheduling for fog nodes and assume that the NSP 
has an unlimited energy supply (e.g., powered by the electric grid). 
Several previous works considered renewable-energy-supported communication services for large-scale network infrastructures\cite{XY2014GC14CoopEnergHarvest,Chia2014RenewableBSs, Li2016RenewableBSs}. How to optimize the utilization of renewable energy for both communications and computational services 
is left for our future work.

\subsubsection{Computational Resource Constraint}
We assume each fog node has limited computational resources that can be dynamically activated and deactivated according to energy availability. Various energy-control approaches have been proposed to allow electronic devices to dynamically adjust their power consumption according to the available energy supply\cite{Zhang2015DataCenter}. For example, a fog node can scale up or down frequencies of its processing units depending on 
the computational workload. Another simpler and more widely adopted approach is to allow each node to dynamically switch on and off some of its 
processing units according to energy availability\cite{Tran2015DataCenter}. In this paper, we adopt the latter approach and assume that each fog node $i$ has a minimum amount of computational resource, measured by the amount of workload that can be processed by each of its processing units per time slot. Let $w_i$ be the service rate that can be provided by each unit of computational resource at fog node $i$.  
%
Let $p^{(k)}_{i,t}$ be the number of computational resource units activated by fog node $i$ to serve the $k$th service type during time slot $t$. The maximum service rate that can be supported by node $i$ for service type $k$ 
can then be written as $w_i p^{(k)}_{i,t}$. The workload that can be offloaded by each node $i$ during time slot $t$ cannot exceed the maximum service rate. In particular, suppose fog node $i$ can only offload $\hat \alpha^{(k)}_{i,t}$ portion of its received type $k$ workload request $\lambda^{(k)}_{i,t}$ for $0 \le \hat \alpha^{(k)}_{i,t} \le 1$. The computational resource constraint at each fog node $i$ is given by:
\begin{eqnarray}
\hat \alpha^{(k)}_{i,t} \lambda^{(k)}_{i,t} \le w_i p^{(k)}_{i,t}, \; \forall i\in {\cal F} \mbox{ and } k\in {\cal V},
\label{eq_Constraint1_CompResource}
\end{eqnarray}
where if $\hat \alpha^{(k)}_{i,t} = 0$, node $i$ cannot process any type $k$ workload received in time slot $t$. In this case, fog node $i$ will have to forward all the received type $k$ workload to other neighboring fog nodes or directly send 
to the CDCs. If $\hat \alpha^{(k)}_{i,t} = 1$, then fog node $i$ will process all the received type $k$ workload. Note that it is unnecessary to force every fog node to distribute resources to process all supported types of services. For example, some fog nodes can allocate all their computational resources to support a limited number of service types and forward the workload associated with other service types to other neighboring fog nodes or CDCs.  


\blu{It is known that the energy-harvesting fog computing networks can exhibit significant temporal and spatial variations. 
The workload processing capability for each fog node can be further improved if it can forward some of its received workload to other neighboring nodes when it cannot harvest sufficient energy, 
and also help some of other nodes to process their workload when it harvests more renewable energy than needed. 
We therefore follow the same line as \cite{XY2017FogCompInfocom} and consider a fog node cooperation strategy, referred to as {\it offload forwarding}, in which two or more fog nodes can help each other and jointly process their workloads. Note that allowing every fog node to always forward part of its workload to all the other fog nodes is uneconomic and difficult to manage. Each fog node should only coordinate with a limited number of the closely located fog nodes 
via the regional SDN-based orchestrator. 
Let $\cC_{i}$ be the set of neighboring fog nodes that can be coordinated with fog node $i$, $\cC_{i} \subseteq {\cal F}\backslash \{i\}$.} 

Suppose fog node $i$ decides to process $\hat \alpha^{(k)}_{i,t} \lambda^{(k)}_{i,t}$ workload of type $k$ service in time slot $t$ with the help of its neighboring fog nodes. Node $i$ will need to carefully divide this total workload into $|\cC_{i}|$ partitions each of which can be forwarded to its neighboring fog nodes. Let $\alpha^{(k)}_{im, t}$ be the portion of type $k$ load received by fog node $i$ to be forwarded to fog node $m$, $m\in {\cC_{i}}$, in time slot $t$. We also use $\alpha^{(k)}_{ii, t}$ to denote the portion of type $k$ load that will be processed by node $i$ itself. 
Clearly,
\begin{eqnarray}
0 < \hat \alpha^{(k)}_{i,t} = \sum_{j\in {\cC_{i}}\cup \{i\}} \alpha^{(k)}_{ij,t} \le 1, \; \forall i\in {\cal F}.
\label{eq_Constraint_CoopOffloading}
\end{eqnarray}



%
\subsubsection{Energy Constraint}
Let ${e}_{i,{\rm unit}}$ be the amount of energy consumed by fog node $i$ to activate each unit of computational resource. We write the total energy consumed by fog node $i$ to process type $k$ service in time slot $t$ as $e^{(k)}_{i,t} = {e}_{i,{\rm unit}} p^{(k)}_{i,t}$. The total energy consumed by fog node $i$ during slot $t$ is given by $e_{i,t} = \sum _{k \in \cV} e^{(k)}_{i,t}$.

Each node $i$ has installed a battery that can store up to ${e}_{i,{\rm max}}$ energy. 
We consider an energy-harvesting system with causality constraints. In particular, a node cannot consume the energy that will be harvested in the future. We further assume that a node cannot use the energy harvested in the current time slot. 
This is because the energy harvested by each fog node can be highly unstable and fluctuated. Most energy harvesting-based electronic devices have an energy-stabilizing circuit to stabilize the energy input into the battery, and each device is typically supplied by the energy output from its battery. 
We can write the battery level of node $i$ at the beginning of time slot $t$ ${\tilde e}_{i,t}$ as
\begin{eqnarray}
{\tilde e}_{i,t} = \min \{ e_{i, {\rm max}}, {\tilde e}_{i,t-1} + \hat e_{i,t-1} - e_{i,t-1}\},
\label{eq_BatteryLevel}
\end{eqnarray}
where $\hat e_{i,t-1}$ is the amount of energy that can be harvested by fog node $i$ during time slot $t-1$.
\blu{We can therefore write the energy constraint of 
each fog node $i$ during time slot $t$ as:
\begin{eqnarray}
\sum _{k \in \cV} e^{(k)}_{i,t} \le {\tilde e}_{i,t},
\label{eq_Constraint2_Energy}
\end{eqnarray}
where if fog node $i$ decides to only process a subset of service types, we have $e^{(k)}_{i,t} = 0$ for some $k \in {\cal V}$.}

\vspace{-0.1in}
\subsection{Problem Formulation}
\label{Subsection_ProblemFormulation}

\blu{In each time slot $t$, fog node $i$ needs to decide the energy allocated to activate the computational resources for each type of services. Fog node $i$ also needs to decide the workload to be processed by itself as well as those to be forwarded to others. In particular, each fog node $i$ needs to decide the following two vectors:}
\begin{itemize}
\item[1)] {\it Energy distribution vector} $\bd_{i,t} = \langle e^{(k)}_{i,t} \rangle_{k\in {\cal V}}$ specifies the amount of energy spent on activating the computational resources to serve different types of services.
\item[2)] {\it Workload offloading vector} $\balpha_{i,t} = \langle \balpha^{(k)}_{i,t} \rangle_{k\in {\cal V}}$ where
\blu{
\begin{eqnarray}
\balpha^{(k)}_{i,t} = \left\{ {\begin{array}{*{20}{l}}
{\hat \alpha^{(k)}_{i,t},}&\mbox{no offload forwarding},\\
{\alpha^{(k)}_{im,t},}&\mbox{with offload forwarding,}
\end{array}} \right. \nonumber
\end{eqnarray}
specifies the portions of received workload to be processed by fog node $i$ with/without the help of its neighboring fog nodes.}   
\end{itemize}

We assume that the 
CDC can offer a certain reward to incentivize the workload offloading behaviors of fog nodes. Each fog node receives reward only for workload that can be processed with the satisfactory QoS measured by the service response-time, i.e., the response-time $\pi^{(k)}_{i,t} \left( \balpha^{(k)} _{i,t} \right)$ 
for type $k$ service of fog node $i$ needs to satisfy $\pi^{(k)}_{i,t}\left( \balpha^{(k)} _{i,t} \right) \le \theta^{(k)}$, $\forall k\in {\cal V}$. The reward received by each node is closely related to the total amount of offloaded workload and types of service.
This reward can be a monetary value paid to each fog node owner (e.g., FSP) or a virtual currency that can be used by fog nodes to exchange a certain service from CDCs. If fog nodes are deployed and managed by the NSP, the reward can be regarded as the control mechanism imposed by the NSP to regulate the workload offloading behaviors of fog nodes.
%
%
%
$\pi^{(k)}_{i,t}\left( \balpha^{(k)} _{i,t} \right)$ depends on the amount of offloaded workload and energy distributed for type $k$ service. 
%
%
%
For example, suppose that fog node $i$ decides to offload $\alpha^{(k)} _{i,t}\lambda^{(k)}_{i,t}$ workload of type $k$ service by itself with $p^{(k)}_{i,t} = {e^{(k)}_{i,t} / {e}_{i, {\rm unit}}}$ activated computational resource units. Consider, for simplicity, an M/M/1 queuing delay for each type of service at fog nodes. The response-time for type $k$ service offloaded by fog node $i$ in time slot $t$ can be written as \blu{(Please see Appendix \ref{AppendixA} for the detailed derivation.)}
        \begin{eqnarray}
       \lefteqn{
        \pi^{(k)}_{i,t} \left( \balpha^{(k)} _{i,t} \right) = 
        \sum\limits_{m\in \cC_{i} \cup \{i\}} \alpha^{(k)}_{im,t} \left( \tau_{im} \right.} \nonumber \\
       &&\;\;\;\;\;\;\;\; \left. + {1 \over  w_m p^{(k)}_{m,t} - \sum_{j \in {\cC}_{i} \cup \{i\}} \alpha^{(k)}_{jm,t} \lambda^{(k)}_{j,t} } \right)  
        \label{eq_ResponseTime1FogNodeOfflForward}
        \end{eqnarray}
where $\tau_{im}$ is the round-trip time between fog nodes $i$ and $m$, with $\tau_{im} > 0$ for $m \neq i$ and $\tau_{ii}=0$.   \blu{Note that $\alpha^{(k)}_{jm,t} \neq 0$  for $m\in {\cC}_j$ means that fog node $m$ has extra computational resource to process the workload sent by fog node $j$.}

Let $\rho^{(k)}_i$ be the reward received by node $i$ to successfully process each workload unit of type $k$. 
Generally speaking, the higher the required QoS, the higher the reward received by the service supporting fog nodes. However, with more stringent QoS, the total amount of workload that can be offloaded by each fog node will also be limited. The main objective for each fog node is to maximize the amount of offloaded workload that can be supported within the tolerable response-time. We follow a commonly adopted setting and assume the reward of fog node $i$ is linearly proportional to the amount of offloaded services. More specifically, the reward obtained by fog node $i$ from offloading type $k$ service is given by

%
\begin{eqnarray}
\varpi^{(k)}_{{i,t}} \left( \balpha^{(k)}_{{i,t}} \right) = \rho^{(k)}_i \sum_{m\in {\cC_{i}}\cup \{i\}}\alpha^{(k)}_{im,t} \lambda^{(k)}_{i,t},
\end{eqnarray}
where $\alpha^{(k)}_{im,t}$ is the portion of workload that can be offloaded within the tolerable response-time.


Motivated by the fact that practical network systems can often tolerate small periods of performance degradation as long as the long-term average performance is good, we consider a system in which the main objective for each fog node is to maximize its long-term discounted reward.  
We can then write the sequential resource slicing and workload offloading problem for each fog node $i$ as follows:

\begin{eqnarray}
&& \max\limits_{\langle \bd_{i,t}, \balpha_{i,t} \rangle_{t=0,1,\ldots}} \; \mathbb{E} \left( \lim\limits_{N \rightarrow \infty} \sum\limits^{N}_{t=0} \gamma^t  \sum\limits_{k\in {\cal V}} \varpi^{(k)}_{{i,t}} \left( \balpha^{(k)}_{{i,t}} \right) \right)
\label{eq_MaxOffload1FogNode}\\
&&\;\;\;\;\; \mbox{s.t.} \;\;\; \pi^{(k)}_{i,t} \left( \balpha^{(k)}_{{i,t}} \right) 
\le \theta^{(k)} \mbox{ and constraints  (\ref{eq_Constraint1_CompResource}), (\ref{eq_Constraint_CoopOffloading}), (\ref{eq_Constraint2_Energy})} \nonumber  \\
&&\;\;\;\;\;\;\;\;\;\;\;\;\;\;\;\;  \forall k\in \cV,  t=0,1,\ldots , \nonumber  
\end{eqnarray}
\blu{where $\gamma$ is the discount factor specifying how
impatient fog node $i$ is, i.e., the smaller the value of $\gamma$, the more node $i$ cares about the reward in the current time slots than the future.}

%
\begin{figure}
\centering
\includegraphics[width=3.2 in]{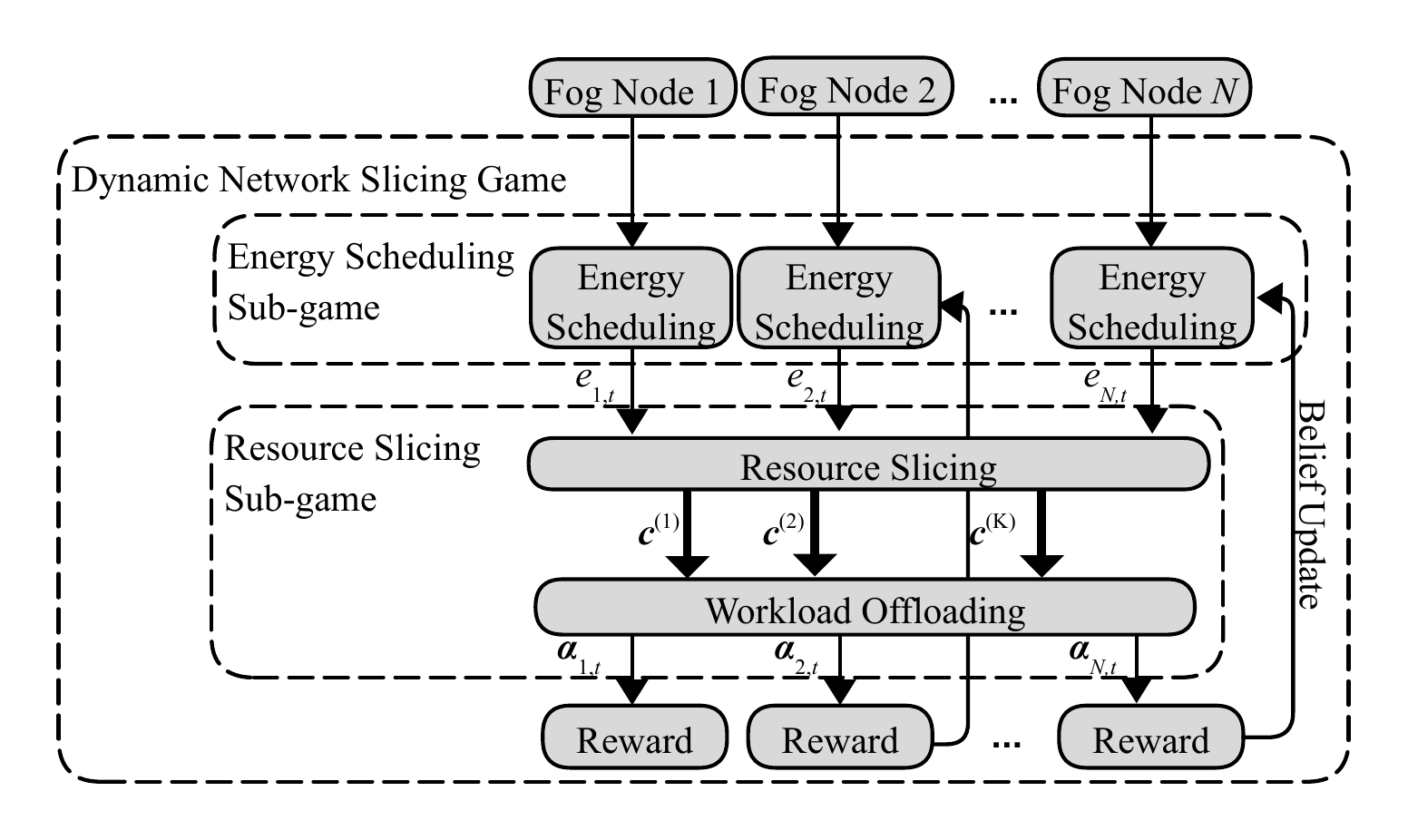}
\vspace{-0.2in}
\caption{Relationship between different sub-games in dynamic\newline network slicing game.}
\label{Figure_DynamicSlicingDiagram}
\vspace{-0.3in}
\end{figure}

\vspace{-0.1in}
\section{Dynamic Network Slicing Game}
\label{Section_Game}

Let us introduce the {\em dynamic network slicing game} 
consisting of two sub-games: resource slicing sub-game and energy scheduling sub-game, as illustrated in Figure \ref{Figure_DynamicSlicingDiagram}.

\subsection{Resource Slicing Sub-game}
\label{Subsection_ResourecDistribution}
We first consider resource distribution among fog nodes in a single time slot $t$. 
Suppose the workload arrival rates $\blambda_{i,t} = \langle \lambda^{(k)}_{i,t} \rangle$ of every fog node $i$ is fixed. Each fog node $i$ has already decided the total amount of energy $e_{i,t}$ it will spend on 
serving the supported service, $0 < e_{i,t} \le \tilde e_{i,t}$. Note that $e_{i,t}$ specifies the total amount of resources that is available at node $i$ to distribute among different types of service in time slot $t$. We use $\be_t = \langle e_{i,t} \rangle_{i\in {\cal F}}$ to denote the energy usage vector for all fog nodes in time slot $t$. We discuss how to schedule energy usage  over different time slots in the next subsection.

The main objective for each fog node $i$ is to carefully decide the energy distribution vector $\bd_{i,t}$ and workload offloading vector $\balpha_{i,t}$ for the rest of the time slot $t$. Each fog node 
should always utilize all the activated energy resource, i.e., $\bd_{i,t}$ needs to satisfy $e_{i,t} = \sum_{k\in {\cal V}} e^{(k)}_{i,t}$.
%
We use the framework of the overlapping-coalition-formation game to model the energy and computational resource slicing/partioning problem among fog nodes. The overlapping coalition formation game has been widely applied to study the resource allocation problem among multiple players. In this game, multiple players can form different coalitions and distribute their resources to serve different types of services. Coalitions are overlapped when the same player joins different coalitions to serve different service types\cite{Chalkiadakis2010OverlapCoalitioanGame}.


Let us formally introduce the resource slicing sub-game. 
\begin{definition}
\label{Definition_OverlapCoalitionForm}
A \emph{resource slicing sub-game} is defined by a tuple ${\cal G} = \langle {\cal F}, {\be}_t, {\cal V}, \bvarpi \rangle$ where ${\cal F}$ is the set of fog nodes that correspond to the players of the game, $\be_t = \langle e_{i,t} \rangle_{i \in {\cal F}}$ is the energy resources that can be distributed by fog nodes, $\cal V$ is the set of service types for each fog node to distribute energy, 
$\bvarpi$ is the vector of rewards received by fog nodes.
\end{definition}

We give a more detailed discussion for each of the above elements as follows. Each fog node can divide its energy $e_{i,t}$ into different partitions (slices) each of which will be allocated to activate the computational resource to support a specific type of service. Each fog node 
needs to carefully decide the amount of workload that can be offloaded by itself and/or with the help of its neighboring fog nodes. 
The main objective for each fog node is to maximize its reward received in the currently time slot.
%
We define a (resource) slice ${\bc^{(k)}}$ for type $k$ service as a vector of energy distributed by fog nodes to serve type $k$ service, i.e., ${\bc^{(k)}} = \langle e^{(k)}_{i,t} \rangle_{i\in \cF}$. Note that it is not necessary for every fog node to distribute energy to support all types of services, e.g., fog node $i$ may have $e^{(k)}_{i,t} = 0$ for some $k\in {\cal V}$. We denote the support of $\bc^{(k)}$ as ${\rm supp} \left( \bc^{(k)} \right) = \{i\in \cF: e^{(k)}_{i,t} \neq 0 \}$.
We define a (resource) slicing structure ${\bc} = \langle \bc^{(k)} \rangle_{k\in {\cal V}}$ as a vector specifying the energy allocation for fog nodes among all supported services.

We consider a {\em transferrable-utility game} setting in which the total reward obtained by a slice can be freely transferred among contributing fog nodes. In this case, the main objective for each fog node is to maximize the total workload offloaded for each type of service. We define the {\em worth} of slice $\bc^{(k)}$ as the total reward that can be obtained by all the member fog nodes distributing energy to type $k$ service. In particular, we write the worth of a slice $\bc^{(k)}$ as
\begin{eqnarray}
v\left(\bc^{(k)}\right) = \sum\limits_{i\in {\rm supp} (\bc^{(k)})} \varpi^{(k)}_{i, t} \left( \alpha^{(k)}_{i,t} \right).
\label{eq_WorthofSlice}
\end{eqnarray}

Let us consider the workload offloading vectors for each given slicing $\bc$. We can show that the optimal workload offloading vector $\balpha^{(k)*}$ for a given slicing structure $\bc$ that maximizes the reward of slice for type $k$ service is unique. More specifically, let us first consider the case that each fog node cannot seek help from other fog nodes but can only process the received workload by its on-board processors. 
Since the reward received by each fog node $i$ is proportional to the amount of offloaded workload, for a given energy distribution vector $\bd_{i,t}$, 
the optimal portion of offloaded workload can be derived by setting $\pi^{(k)}_{i,t} \left( \hat \alpha^{(k)}_{i,t} \right)=\theta^{(k)}$ (See (\ref{eq_ResponseTime1FogNodeNoCloud}) in Appendix A for the expression of $\pi^{(k)}_{i,t}$) which is given by
\begin{eqnarray}
\alpha^{(k)*}_{i,t} \left( e^{(k)}_{i,t} \right) = \min\left\{1,  {w_i e^{(k)}_{i,t} \over \lambda^{(k)}_{i,t} e_{i,{\rm unit}}} - {1 \over \theta^{(k)} \lambda^{(k)}_{i,t}} \right\} \; \forall k\in {\cal V}.
\label{eq_OptimalAlpha_LocalOffload}
\end{eqnarray}

From (\ref{eq_OptimalAlpha_LocalOffload}), we can observe that since, for a given energy distribution vector $\bd_{i,t} = \langle e^{(k)}_{i,t} \rangle_{k\in {\cal V}}$, the optimal workload offloading vector $\balpha^{*}_{i,t} \left( \bd_{i,t} \right) = \langle \alpha^{(k)*}_{i,t} ( e^{(k)}_{i,t} ) \rangle_{k\in {\cal V}}$ can be fully determined,  the joint optimization problem for both computational and energy resource distribution is equivalent to only optimizing the energy distribution $\bd_{i,t}$ among all supported types of service.

If each fog node can use offload forwarding to cooperate with its neighboring fog nodes as described in Section \ref{Section_SystemModel}, we can then write the optimal workload offloading vector $\balpha^{(k)*} \left( \bc^{(k)} \right) = \langle \balpha^{(k)*}_{i,t} \left( \bc^{(k)} \right) \rangle_{k\in {\cal V}, m\in {\cal C}\cup \{i\}, i\in {\cal F}}$ where 
\begin{eqnarray}
\lefteqn{ \balpha^{(k)*}_{i,t} \left( \bc^{(k)} \right) = \arg \max\limits_{\balpha^{(k)}_{i,t}
} \varpi^{(k)}_{i,t} \left( \balpha^{(k)}_{i,t} ( \bc^{(k)} ) \right)} \label{eq_MaxOffloadMultipleFogNodes_SlicingSubgame} \\ 
&&\mbox{s.t.}\;\;\;  \pi^{(k)}_{i,t} \left( \balpha^{(k)}_{i,t}\right) \le \theta^{(k)} \mbox{ and constraints (\ref{eq_Constraint1_CompResource}), (\ref{eq_Constraint_CoopOffloading}),(\ref{eq_Constraint2_Energy})}, \nonumber \\
&&\;\;\;\;\;\;\;\; \forall k \in \cV, t=1,2,\ldots.
\end{eqnarray}


%
Note that since $\varpi^{(k)}_{i,t} \left( \balpha^{(k)}_{i,t} \right)$ can be considered as a linear combination of $\alpha^{(k)}_{im,t}$ with different weights $\lambda^{(k)}_{i,t}$, optimal solution $\balpha^{(k)*}_{i,t} \left( \bc^{(k)} \right)$ is unique for the given $\bc^{(k)}$. In other words, instead of optimizing both $\balpha^{(k)*}_{i,t}$ and $\bc^{(k)}$, fog nodes only needs to decide the optimal slicing $\bc^{(k)*}$ for each service type $k$.

We can observe that the worth function is monotone, i.e., $v\left(\bc^{(k)}\right) \ge v\left(\bc^{(k)'}\right)$ for any $\bc^{(k)}$, $\bc^{(k)'}$ such that $e^{(k)}_{i,t} \ge e^{(k)'}_{i,t}$ for all $i\in {\cal F}$. In other words, the more energy has been distributed to a slice, the higher reward can be received by the fog nodes. 

We define a reward allocation among member fog nodes for each slice $\bc^{(k)}$ as $\bvarpi^{(k)} = \langle \varpi^{(k)}_{i, t} \rangle_{i \in {\rm supp} (\bc^{(k)})}$ which describes the worth distributed among fog nodes for serving type $k$ service. $\bvarpi^{(k)}$ is said to be {\em efficient} if $\sum_{i\in {\rm supp} (\bc^{(k)})} \varpi^{(k)}_{i,t} = v\left(\bc^{(k)}\right)$. $\bvarpi^{(k)}$ is also called {\em imputation} if it is efficient and satisfies the individual rationality, i.e., $\varpi^{(k)}_{i,t} \ge v \left( \underline{\varpi}^{(k)}_{i, t} \right)$ where $\underline{\varpi}^{(k)}_{i, t}$ is the reward obtained by fog node $i$ for offloading type $k$ service if fog node $i$ cannot cooperate with other fog nodes.
We refer to a (resource) slicing agreement as a tuple $\langle \bc, \bvarpi \rangle$ where $\bc = \langle \bc^{(k)} \rangle_{k\in {\cal V}}$ and $\bvarpi = \langle \bvarpi^{(k)} \rangle_{k\in {\cal V}}$.

The resource distribution and negotiation among fog nodes across different types of services can be very complex. For example, when two or more fog nodes decide to cooperate to offload a specific type of service, they can also impose a certain term that may affect their cooperation with other fog nodes when serving some other types of service. When a fog node deviates from a slicing agreement for a specific type of service, it will also affect its cooperation with other fog nodes in other service types. The main solution concept in the resource slicing sub-game is the {\em core}.
We extend the concept of the conservative core in the overlapping coalition formation game into our resource slicing sub-game as follows:
\begin{definition}
Given a resource slicing sub-game ${\cal G} = \langle {\cal F}, \bc_t, {\cal V}, {\bvarpi} \rangle$ and a subset of fog nodes ${\cal N} \subseteq {\cal F}$.  Suppose $\langle \bc, \bvarpi \rangle$ and $\langle \bc', \bvarpi' \rangle$ are two slicing agreements such that for any slice $\bc^{(k)} \in \bc$ either ${\rm supp} (\bc^{(k)}) \subseteq {\cal N}$ or ${\rm supp} (\bc^{(k)}) \subseteq {\cal F}\setminus {\cal N}$. We say that slicing agreement $\langle \bc', \bvarpi' \rangle$ is a profitable deviation of ${\cal N}$ from $\langle \bc, \bvarpi \rangle$ if for all $j \in {\cal N}$, we have $\varpi_{j} \left( \bc' \right) > \varpi_{j} \left( \bc \right)$. We say that a slicing agreement $\langle \bc, \bvarpi \rangle$ is in the core of $\cal G$ if no subset of $\cal F$ has a profitable deviation from it. In other words, for any subsets of fog nodes ${\cal N} \subseteq {\cal F}$, any slicing structure $\bc'$, and any imputation $\bvarpi'$, we have $\varpi'_{j} \left( \bc' \right) \le \varpi_{j} \left( \bc \right)$.
\end{definition}
%
We can prove the following result. 
\begin{theorem}
\label{Theorem_Optimal}
The core of the resource slicing sub-game is non-empty and consists of an unique outcome that maximizes the social welfare.
\end{theorem}
\begin{IEEEproof}
See Appendix \ref{Proof_Optimal}.
\end{IEEEproof}

From Theorem \ref{Theorem_Optimal}, we can observe that if all the fog nodes have decided their total amount of energy consumed in each time slot, the optimal slicing structure that maximizes the total reward of the fog nodes is unique and stable. 
In the next subsection, we investigate the energy scheduling for fog nodes in which each fog node $i$ seeks an optimal energy scheduling policy that can maximize its long-term discounted reward.


\subsection{Energy Scheduling Sub-game}

From the previous discussion, we can observe that resource slicing among fog nodes in each time slot is closely related to the total amount of energy that can be used by fog nodes to activate their computational resources. Each fog node can further improve its long-term reward by scheduling its energy usage over different time slots according to the evolution of the energy harvesting and workload arrival processes. In this section, we assume for each given energy distribution vector $\be_{t}$, fog nodes can always partition their energy among all the supported types of service as described in Section \ref{Subsection_ResourecDistribution}. We model the scheduling of energy usage among fog nodes
in a time-varying environment as a stochastic game, referred to as the energy scheduling sub-game. 

\begin{definition}
A {\em energy scheduling sub-game} is a tuple ${\cal G}' = \langle {\cF}, \bcY, \bcA, \cS, \Omega, \bTheta, T, \bvarpi, \rangle$ where ${\cF}$ is the set of fog nodes, $\bcY = \times_{i\in \cF} \cY_i$ is the set of type profile where $\cY_i$ is the type space for each fog node $i$, $\bcA$ is the set of action profile, $\cS$ is the set of possible outcome states, $\Omega$ is the set of observations that can be obtained by each fog node, $\bTheta$ is the observation function of fog nodes, $T\left( s', \ba, s \right)$ is the state transition function, and $\bvarpi$ is the vector of reward functions for fog nodes.
\end{definition}

We give a detailed discussion on each of the above elements as follows: each fog node (player) is assumed to be rational and always tries to maximize its long-term discounted reward. The {\em action} of each fog node corresponds to the energy it scheduled  on activating the computational resources as well as the resource distribution in each time slot, i.e., we have $a_{i,t} = \langle \bd_{i,t}, \balpha_{i,t}, \be_{i,t} \rangle$ for $\balpha_{t} = \langle \alpha_{i,t} \rangle_{i\in {\cal F}} \in {\bcA}$. 
%
The {\em type} of each fog node reflects its ``workload offloading capability" related to how much this fog node can help or rely on other fog nodes to offload service workload. It depends on the battery level, computational power, and the number of neighboring fog nodes, QoS requirements of associated users, as well as the long-term energy harvesting and workload arrival processes. Each fog node can perfectly know its own type but not those of others. The {\em state}, denoted as $s_t \in \cS$ in time slot $t$, is a composite variable of the battery level $\tilde \be_t$ and workload arrival rates $\blambda_t$ of fog nodes. It is known that if the duration of each time slot is short enough, the energy harvesting and workload arrival processes for fog nodes can possess the Markov property, that is the conditional probability distribution of future states only depends on the present state\cite{Aprem2013EnergHarv,XY2015GlobecomEnergHarvest}. Let $\Pr\left( \hat \be_{t}| \hat \be_{t-1} \right)$ and $\Pr\left( \blambda_{t} | \blambda_{t-1} \right)$ be the transition probabilities for energy harvesting and workload arrival process between two consecutive time slots where $\hat \be_{t} = \langle \hat e_{i,t} \rangle_{i\in {\cal F}}$ and $\blambda_{t} = \langle \lambda_{i,t} \rangle_{i\in {\cal F}}$. Since the workload arrival and energy harvesting are two independent processes, we can then apply Bayes' rule\cite{Chalkiadakis2010SequentialCoalitioanGame, XY2015GlobecomEnergHarvest} and calculate the state transition function $T \left( s_{t}, s_{t-1}, \ba_{t-1} \right)$ as follows:
\begin{eqnarray}
\lefteqn{ T \left( s_{t}, s_{t-1}, \ba_{t-1} \right) = \Pr \left( {\tilde \be}_{t}, \blambda_{t} | \be_{t-1}, \balpha_{t-1}, {\tilde e}_{t-1}, \blambda_{t-1} \right) } \nonumber \\
&&\;\;\;\;\;= \frac{\begin{array}{l}
\Pr \left( \blambda_{t} | \blambda_{t-1} \right)  \Pr \left( {\hat e}_{t} | {\hat e}_{t-1} \right) \\
\;\;\;\;\; \Pr \left( {\tilde e}_{t}, \blambda_{t} | \be_{t-1}, \balpha_{i,t-1}, {\tilde e}_{t-1}, \blambda_{t-1} \right)
\end{array}}{\begin{array}{l}
{\sum_{\blambda_{t}, {\hat e}_{t}} \Pr \left( \blambda_{t} | \blambda_{t-1} \right)  \Pr \left( {\hat e}_{t} | {\hat e}_{t-1} \right)} \\
\;\;\;\;\;{\Pr \left( {\tilde e}_{t}, \blambda_{t} | \be_{t-1}, \balpha_{t-1}, {\tilde e}_{t-1}, \blambda_{t-1} \right)}
\end{array}}
\end{eqnarray}
where $\tilde \be_{t} = \{\tilde e_{i,t}\}_{i\in {\cal F}}$  and $\tilde e_{i,t}$ is given in (\ref{eq_BatteryLevel}).
%
%
Each fog node cannot observe the complete state (e.g., the battery levels and the workload arrival rates of other nodes) at the beginning of each time slot but can obtain a partial observation labeled as $o_{i,t}$. The observation of each fog node at the beginning of each time slot can be its observed initial environmental condition and workload received by itself as well as that from the neighboring fog nodes. For example, if all fog nodes are co-located in the same area, they may observe the similar energy harvesting process (e.g., solar or wind powers collected by co-located fog nodes can have a strong correlation.) and traffic arrival rates (e.g., users located in the same coverage area can have similar traffic patterns.). Each fog node $i$ can then infer the unknown information of other fog nodes from its own observed battery level $\tilde e_{i,t}$ and workload arrival rate ${\lambda}_{i,t-1}$ during the previous time slots. More specifically, fog node $i$ can infer an observation function specifying the probability distribution of all its possible observations given the action profile and resulting state as follows:
\begin{eqnarray}
\lefteqn{ \Theta_{i,t}\left( o_{i,t}, \ba_{i,t-1}, s_{t} \right) = \Pr \left( o_{i,t} | \ba_{i,t-1}, s_{t} \right) } \nonumber \\
&=& {\Pr \left( \tilde \be_{t}, \blambda_{t-1} | \be_{i,t-1}, \balpha_{i,t-1} \right) \over \sum_{\tilde \be_{-i,t}, \blambda_{-i,t}} \Pr \left( \tilde \be_{t}, \blambda_{t-1} | \be_{i,t-1}, \balpha_{i,t-1} \right)},
\end{eqnarray}
where $-i$ denotes all the fog nodes except fog node $i$.
%
%
%
%
Each fog node $i$ can establish a belief function $\hat b_{i} \left( \by_{-i} | s \right)$ about the unknown types of others under each possible state $s \in {\cal S}$. This belief function will help the fog node choose the most ``capable" fog nodes from its neighbors to forward its workload without having the complete information about their types.
Each fog node can decide its action at the beginning of each time slot by utilizing its observation and belief function. The action profile of all the fog nodes will jointly determine  a stochastic outcome state of the game. 
Each outcome state and action profile determine the reward for each fog node.
The main objective for each fog node is to find the optimal policy such that no fog node can further improve its long-term reward by unilaterally deviating from this policy.

Note that since each fog node cannot know the state and types of other fog nodes, it cannot know the exactly value of the final reward obtained when joining each possible coalition.   
%
However, the repeated interaction among fog nodes provides each fog node with the opportunities to learn from the past experience and estimate the expected  reward when it forms different slicing structures with different neighboring fog nodes. More specifically, each fog node $i$ can establish and maintain a belief function reflecting its private belief about the types of neighboring fog nodes. Each fog node $i$'s belief function $b_i$ is a probability distribution function over types of other fog nodes under each given state. We write $b_i \left(\by_{-i}, s\right)$ as the probability that fog node $i$ assigns to other fog nodes in state $s$. 
We follow the commonly adopted setting and assume the decision of each fog node cannot directly affect the decision making process of other fog nodes\cite{Gmytrasiewicz2005IPOMDPs, Chalkiadakis2010SequentialCoalitioanGame}.
%
In addition, each fog node can also utilize its observation to establish a state belief function, a probability distribution function about the possible state, and combine this state belief function with the belief function about the workload offloading capabilities of other fog nodes under each given state. In particular, if a fog node $i$ can have a belief function $b_{i,t} \left( \by_{-i,t}, s_{t} \right)$ about the state and types of other fog nodes at the beginning of time slot $t$, it can calculate the 
%
expected reward $\bar \varpi_{i,t}$ as 
\begin{eqnarray}
\bar \varpi_{i,t} = \gamma \sum\limits_{\begin{subarray}{c} \by_{-i}\in {\boldsymbol \cY}^{|\cC_{i}|-1} \\
s_t\in \cS \end{subarray}}  b_{i,t} \left( \by_{-i,t}, s_{t} \right)
\label{eq_InstantExpectedRewardC}
\sum_{k\in {\cal V}}\varpi^{(k)}_{i} \left( \balpha^{(k)*}_t ( \be^{(k)}_{t}  ) \right).  \nonumber
\end{eqnarray}

It can be observed that if the belief function of each fog node can refect the true state and offloading capability of other fog nodes, each fog node can always decide the optimal actions to maximize its long-term expected rewards.
Let us now introduce a distributed algorithm based on B-POMDP for each fog node to establish and update its belief function about the state and the types of others from its past experience. Each fog node can then use the belief functions to sequentially update its actions to maximize its long-term discounted reward.
We model the decision making process of each fog node as a B-POMDP. B-POMDP extends the traditional single-agent POMDP into multi-agent cases by allowing each agent to include its interactions with other agents as a part of the state space. We present the formal definition as follows:  
\begin{definition}
A belief-state POMDP (B-POMDP) for the decision making process of fog node $i$ is defined as $\mbox{\it B-POMDP}_i = \langle {\cS}_i, {\bcA}, T'_i, b_i, \Omega, \Theta, \varpi_i \rangle$ where ${\cS}_i$ is the state space of fog node $i$, $\bcA$ is the set of action profiles for the fog nodes, $b_i$ is fog node $i$'s belief function which specifies its belief about the state and types of other fog nodes, $T'_i\left( s', \ba, s \right)$ is the transition dynamics specifying the probability of possible outcome state $s'$ for fog node $i$ given that action $\ba$ has been taken in state $s$, $\Omega$ is the set of observation for each fog node, $\Theta$ is the observation function and $\varpi_i$ is the reward function.
\end{definition}

As mentioned earlier, the main difference between the B-POMDP and the single-agent POMDP is that in the former model, each fog node includes the decision making process of other fog nodes as a part of the environmental state. More specifically, in B-POMDP, the state space ${\cS}_i=\cU\times_{j\in \cF\setminus\{i\}}\cM_j$ for each fog node $i$ consists of two parts: the set of states about the physical environment $\cU$ which includes the harvested energy, battery level and received workload, and the set of possible model states about other fog nodes $\times_{j\in \cF\setminus\{i\}}\cM_j$ such as their types that specify the decision making of these fog nodes under each state of physical environment. Each fog node believes that all the other fog nodes decide their actions according to an unknown distribution and fog node $i$'s prior belief about this distribution can follow the Dirichlet distribution\cite{Chalkiadakis2010SequentialCoalitioanGame, Gmytrasiewicz2005IPOMDPs, XY2015DET}.

At the beginning of each time slot, each fog node can obtain an observation $o_{i,t}$ about the current physical environmental state $u_{i,t} \in {\cU}$ and can therefore follow the similar approach as POMDP to updates its belief about the environmental state as follows:
\begin{eqnarray}
\lefteqn{ b'_{i,t} \left( u_{i,t} \right) = \Pr\left( u_{i,t} | o_{i,t}, a_{i,t-1}, {b'}_{i,t-1} \right) }
\nonumber \\
&=& 
       {\Pr\left( o_{i,t+1} | u_{i,t+1}, a_{i, t}, b'_{i,t} \right) \sum_{u_{i,t}\in\cU} \Pr \left( u_{i,t+1} | a_{i, t}, b'_{i,t}, u_{i,t} \right)} \nonumber \\
      && {\Pr\left( u_{i,t} | a_{i, t}, b'_{i,t} \right)}/{\Pr \left( o_{i,t+1} | a_{i, t}, b'_{i,t} \right)} \nonumber \\
&=& \beta \Theta\left( u_{i,t+1}, a_{i, t}, o_{i,t+1} \right) \nonumber \\
&& \sum_{u_{i,t}\in \cU} T \left( u_{i,t}, \ba_{i, t-1}, u_{i,t-1} \right) b'_{i,t-1} \left( u_{i,t-1} \right),
\label{eq_PhyStateBeliefUpdate}
\end{eqnarray}
where $\beta = 1/{\Pr ( o_{i,t+1} | a_{i, t}, b'_{i,t} )}$ is a normalizing factor that is independent of $u_{t}$.

Each fog node cannot observe the types of other fog nodes but can derive the model information about the types of others after it finishes interacting with other fog nodes and receives the reward at the end of each time slot. In other words, there is a mapping function $g_i$ that maps the types of other fog nodes and the physical environmental state of the system to the final slicing structure and reward of fog node $i$, i.e., $\langle \varpi_{i,t}, \bc_{i,t} \rangle = g_i (y_{i,t}, a_{i,t}, \by_{-i,t}, u_t)$ where $\bc_{i,t}$ is the slices that consists of fog node $i$ in time slot $t$. Note that fog node $i$ cannot know $g_i$, but can develop a belief function about other fog nodes' type information by estimating the probability distribution about $\by_{-i,t}$ from its previous observations of $\varpi_{i,t}, \bc_{i,t}, a_{i}, \by_{-i,t}$ and $u_{i,t}$. 
In this paper, we adapt Bayesian reinforcement learning approach for each fog node to learning the unknown type information of other fog nodes\cite{XY2015ICCBayeRL,Chalkiadakis2010SequentialCoalitioanGame}. 
Each fog node can update its belief information about the type of other fog nodes using the observed reward and slicing structure at the end of each time slot via Bayes' rule. More specifically, each fog node $i$ in slices $\bc_{i,t}$ can update its belief function about types of other fog nodes in $\bc_{i,t}$ by
\begin{eqnarray}
\lefteqn{ b''_{i,\bc_{i,t}} \left( \by_{\bc_{i,t}} \right) = \Pr\left( \by_{\bc_{i,t}} | a_{i,t}, u_{t-1}, b''_{i,\bc_{i,t-1}}, o_{i,t}, u_t \right)} \label{eq_TypeBeliefUpdate} \\
&=& {\Pr \left( u_t | a_{i,t}, \by_{\bc_{i,t-1}}, o_{i,t}, b''_{i,\bc_{i,t}} \right) b''_{i,\bc_{i,t-1}} \left( \by_{\bc_{i,t-1}} \right) \over {\Pr \left( u_t | a_{i,t}, o_{i,t} \right)}} \nonumber \\
&=& \beta' T'_i\left( u_t, a_{i,t}, u_{t-1}, \by_{\bc_{i,t-1}} \right) b''_{i,\bc_{i,t-1}} \left( \by_{\bc_{i,t-1}} \right) \nonumber
\end{eqnarray}
where $\beta'$ 
is a normalizing factor that is independent of $\by_{\bc_{i,t}}$. 

Each fog node can then combine the obtained belief functions about the physical environmental state and types of other fog nodes to calculate the expected reward in each time slot. Each fog node will also evaluate the future expected reward by calculating a future updated belief state, i.e., we use $B_i \left( s_i, \ba \right)$ to denote the estimated future belief state when the current state and action of fog node $i$ are given by $s_i$ and $\ba$, respectively.
At the beginning of each time slot $t$, fog node $i$ can calculate ${{\bar \varpi}'}_i \left(\bc_{i,t}, s_t, \ba_{t}, b_{i,t} \right)$ by
\begin{eqnarray}
\lefteqn{{{\bar \varpi}'}_i \left(\bc_{i,t}, s_t, a_{\cC_t}, b_{i,t} \right)= \sum\limits_{\begin{subarray}{c} \by_{\bc_{-i,t}}\in {\boldsymbol \cY}^{|\bc_{i,t}|-1} \\
s_t\in \cS \end{subarray}} b'_{i,t}\left( u_{i,t} \right) b''_{i, \bc_{i,t}}\left( \by_{\bc_{-i, t}} \right) \bar \varpi_{i, t} }\nonumber\\ 
&&\;\;\;\;\; + \gamma \sum\limits_{o_{i,t} \in \Omega} \Theta\left( o_{i,t}, u_{t}, a_{i,t-1} \right) \upsilon_{i} \left( B_{i} \left( s_{i,t}, a_{i, t} \right) \right),
\label{eq_LongTermPayoff}
\end{eqnarray}
where the first term on the right-hand-side of above equation is the expected reward fog node $i$ obtained in the current time slot $t$, and the second term is the expected reward that fog node $i$ can obtain in the future time slots, and $\upsilon_{i} \left( B_{i,t} \left( s_{i,t}, a_{i, t} \right) \right)$ is given by
\begin{eqnarray}
\lefteqn{ \upsilon_{i} \left( B_{i,t} \left( s_{t}, a_{i, t} \right) \right) = \sum\limits_{\bc_{i,t}} \Pr\left( \bc_{i,t}, \ba_{{t+1}}, s_{t+1} | \right.} \\
 &&\;\; \left. B_{i} \left( s_{t}, a_{i, t} \right) \right) {{\bar \varpi}}'_i \left( \bc_{i,t}, s_{t+1}, \ba_{{t+1}}, B_{i} \left( s_{i,t+1}, a_{i, t+1} \right) \right). \nonumber
\end{eqnarray}

From the above analysis, we can write the optimal policy for each fog node to decide its action as:
\begin{eqnarray}
a^*_{i,t} = \max\limits_{a_{i,t} \in \cA} {{\bar \varpi}}'_i \left( \bc_{i,t}, s_t, \ba_{t}, b_{i,t} \right).
\label{eq_OptimalPolicy}
\end{eqnarray}

We can prove the following result.
\begin{theorem}
\label{Theorem_MainAlgorithm}
The belief function in (\ref{eq_TypeBeliefUpdate}) can always converge to a stationary distribution. The policy in (\ref{eq_OptimalPolicy}) is optimal for every initial state. 
\end{theorem}
\begin{IEEEproof}
See Appendix \ref{Proof_Theorm_MainAlgorithm}.
\end{IEEEproof}


\section{Implementation and Numerical Results}
\label{Section_NumericalResults}

\subsection{Dynamic Network Slicing in 5G Networks}
The dynamic network slicing among multiple fog nodes can be supported by the network sharing management architecture recently introduced by 3GPP\cite{3GPP2016NetworkShare, 3GPP2016NetworkShare2}. In particular, a {\em network slice} consists of a set of isolated computational and networking resources (e.g., processing units, network infrastructure, and bandwidth) orchestrated according to a specific type of service. Popular wireless services that can be supported by fog computing include voice processing (e.g., voice recognition applications such Apple's Siri and Amazon's Alexa services) and image processing (e.g., image recognition applications such as the traffic sign recognition in automotive devices).  To facilitate on-demand resource allocation, admission control, workload distribution and monitoring, the regional SDN-based orchestrator can be deployed at the evolved packet core (EPC) with accessibility to the core network elements such as Mobility Management Entity (MME) and Packet Data Gateways (P-GW) via the S1 interface. Each fog node can be deployed inside of the eNB. Regional SDN-based orchestrator can coordinate the workload distribution among connected fog nodes via the X2 interface.
%
Regional SDN-based orchestrator can connect with the network element manager (NEM) of each eNB to evaluate the received workload for every type of service and decide the amount of workload to be offloaded by each connected fog node. It also performs workload monitoring, information exchange and control of the computational resources belonging to different fog nodes through the NEMs of their associated eNBs. 

%

\begin{figure*}
\begin{minipage}[t]{0.37\linewidth}
\centering
\includegraphics[width=2.7 in]{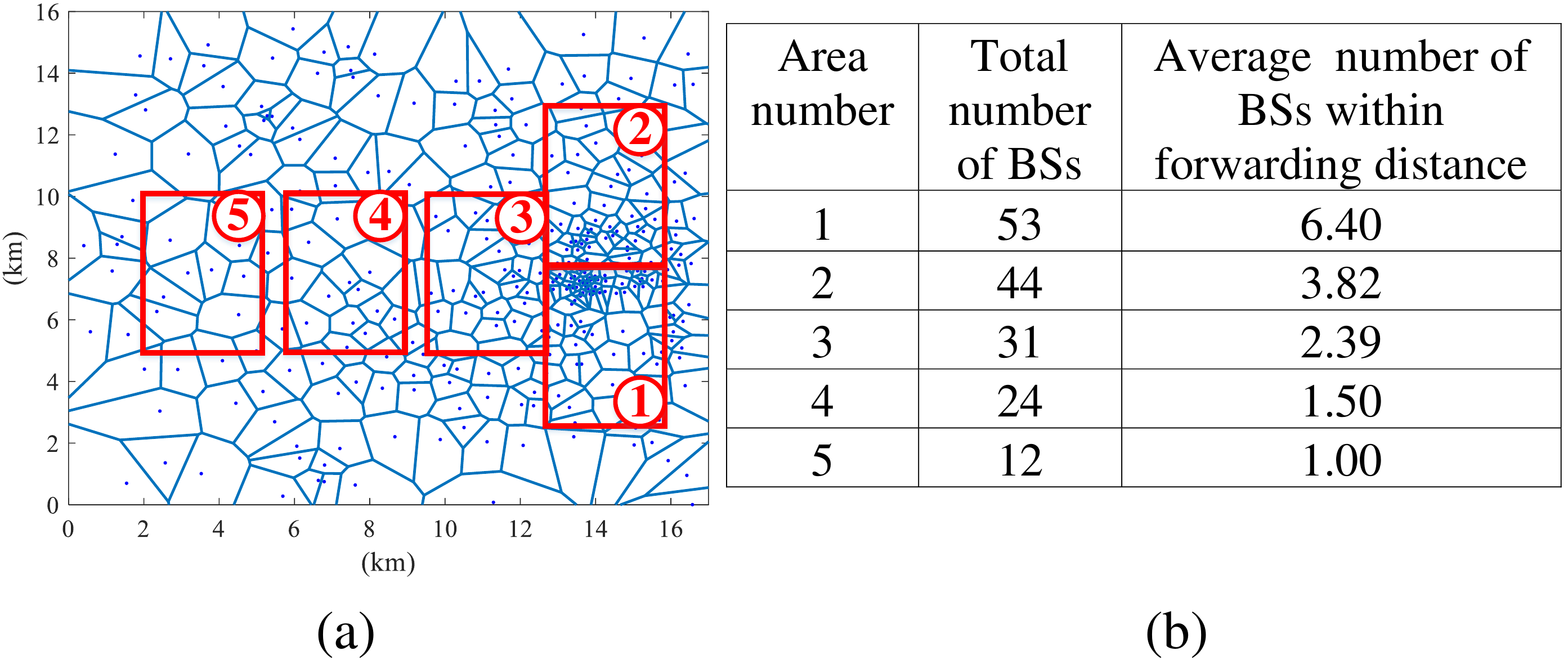}
\vspace{-0.3 in}
\caption{Distribution of BSs and considered areas.}
\label{Figure_BSdistribution}
\vspace{-0.1in}
\end{minipage}
\begin{minipage}[t]{0.2\linewidth}
\centering
\vspace{-1.2in}
\includegraphics[width=1.6 in]{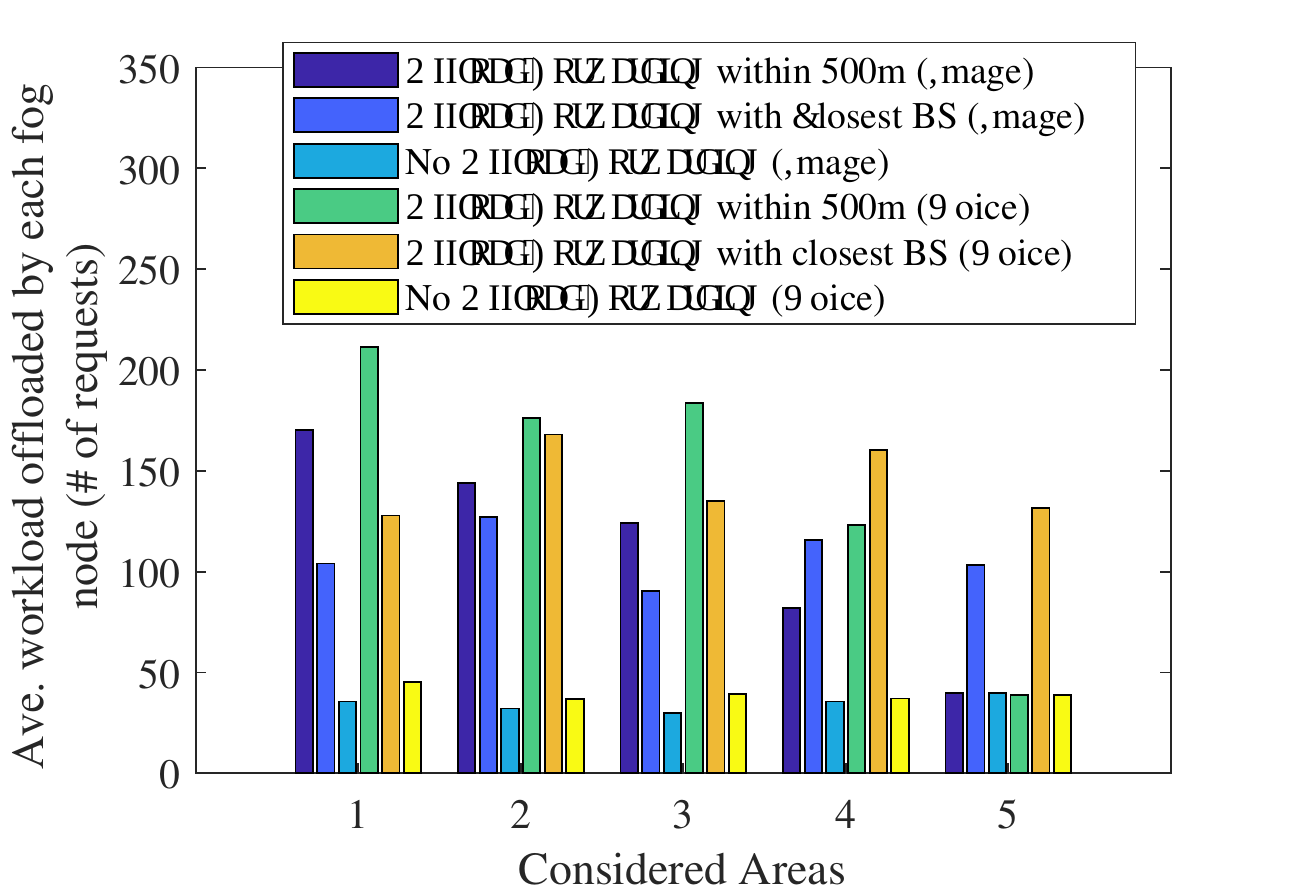}
\vspace{-0.2in}
\caption{Offloaded workload for both types of service in different areas in Figure \ref{Figure_BSdistribution}.} 
\label{Figure_OffloadedWorkloadVSNumberofAreas}
\end{minipage}
\begin{minipage}[t]{0.2\linewidth}
\centering
\vspace{-1.2in}
\includegraphics[width=1.4 in]{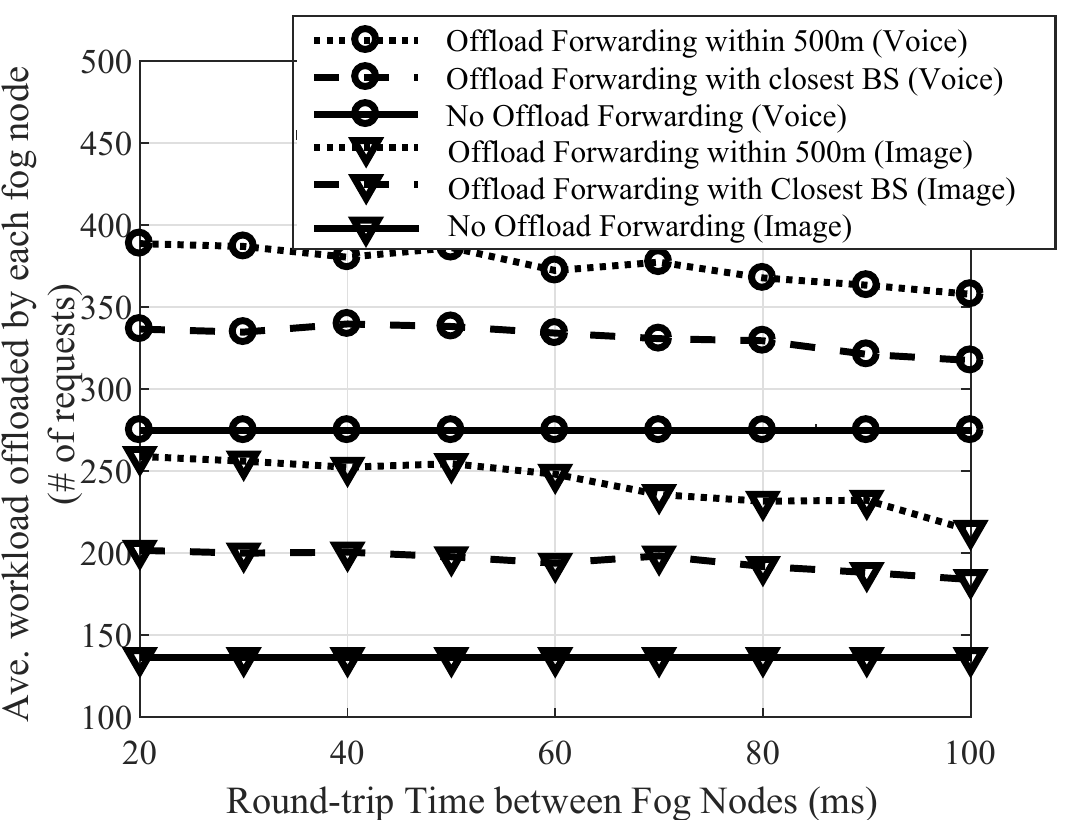}
\vspace{-0.2in}
\caption{Offloaded workload\newline under different RTT between fog nodes.}
\label{Figure_FigOffloadWorkloadVSRoundTripTime}
\end{minipage}
\begin{minipage}[t]{0.2\linewidth}
\centering
\includegraphics[width=1.6 in]{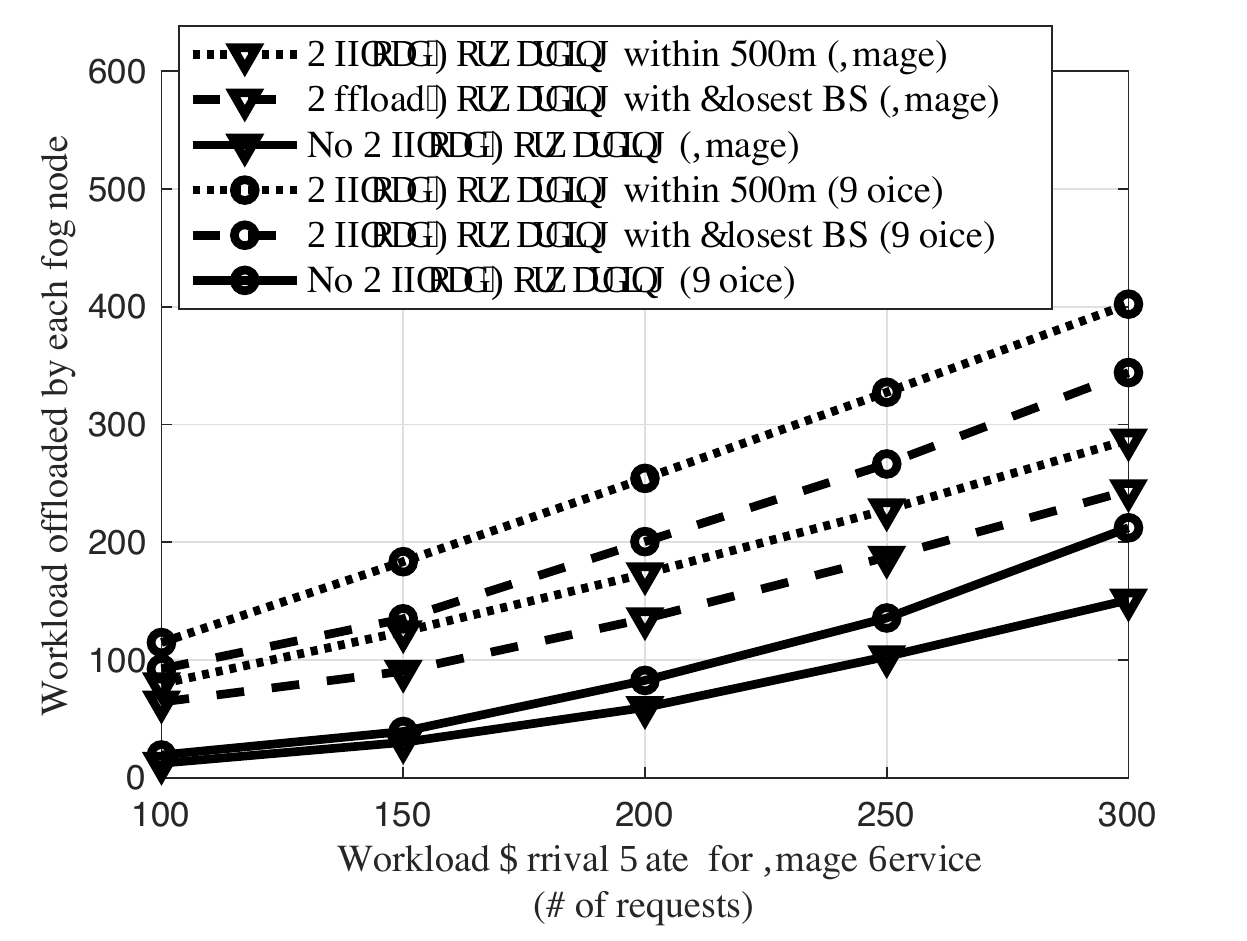}
\vspace{-0.3in}
\caption{Offloaded workload under different workload arrival rate for image service.}
\label{Figure_FigOffloadWorkloadVSArrivalRate}
\end{minipage}
\vspace{-0.3in}
\end{figure*}
\begin{figure}
\begin{minipage}[t]{0.48\linewidth}
\centering
\includegraphics[width=1.7 in]{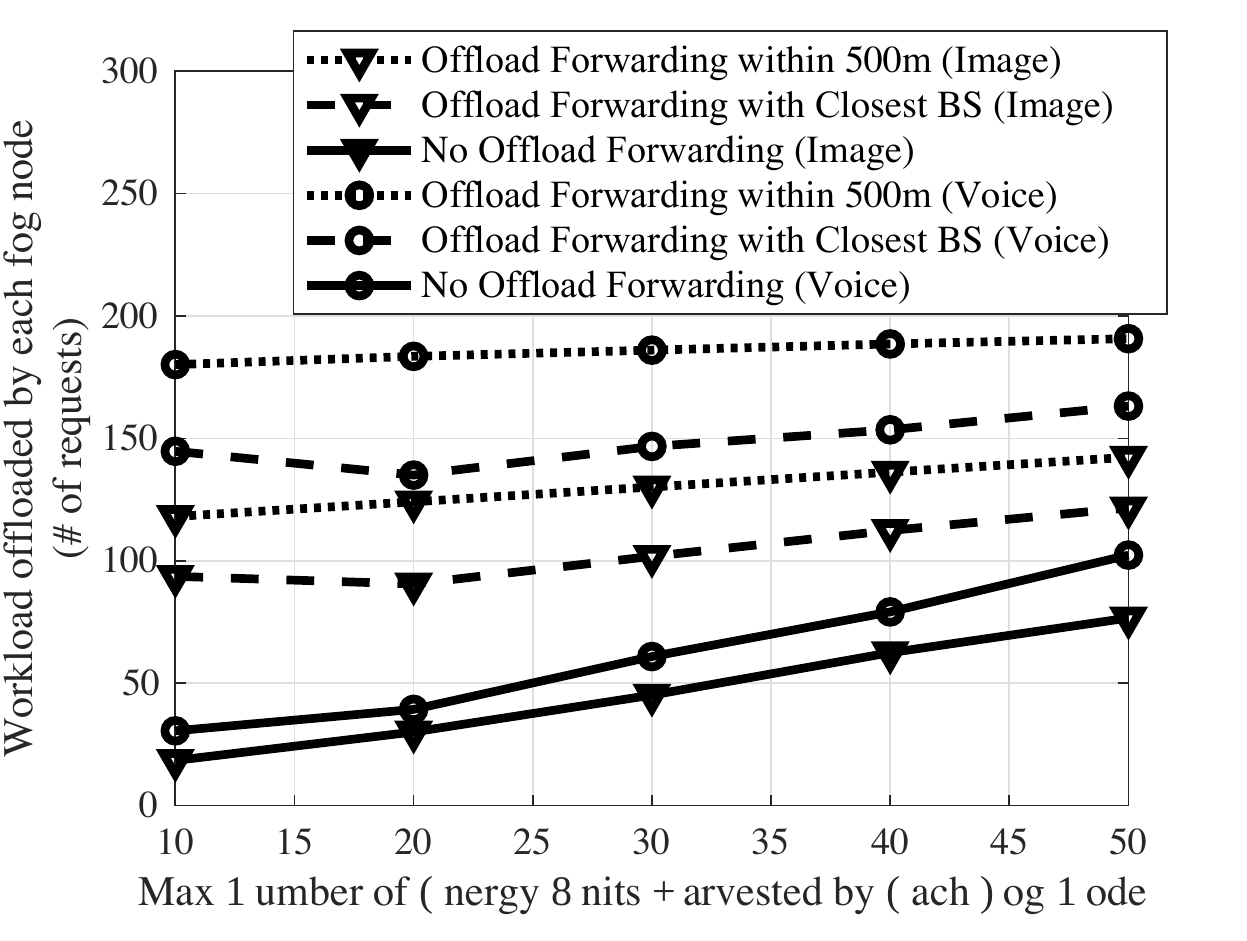}
\vspace{-0.25in}
\caption{Offloaded workload under different amount of harvested energy.}
\label{Figure_FigOffloadVSHarvestEnergy}
\end{minipage}
%
\begin{minipage}[t]{0.5\linewidth}
\centering
\includegraphics[width=1.8 in]{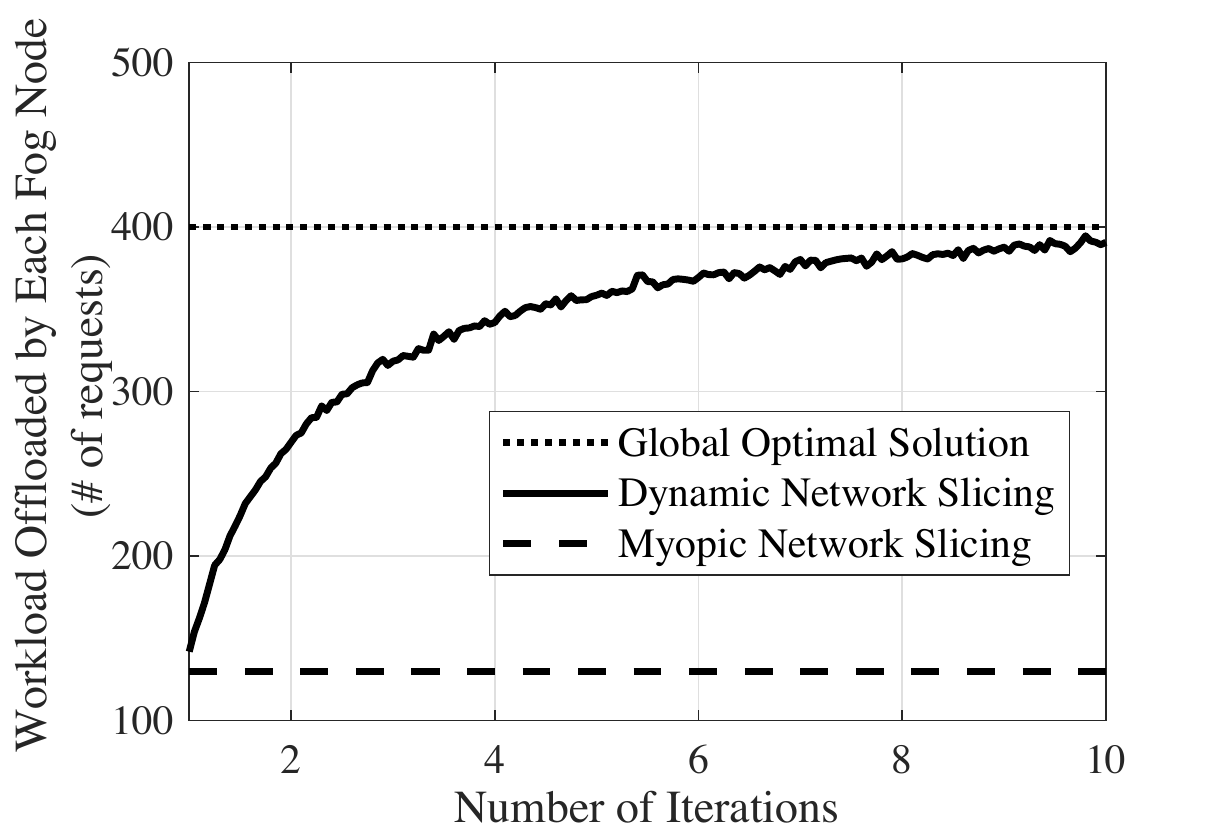}
\vspace{-0.25in}
\caption{Offloaded workload under different numbers of iterations.}
\label{Figure_FigOffloadVSIterations}
\end{minipage}
\vspace{-0.3in}
\end{figure}


\subsection{Numerical Results}
We simulate the possible implementation of fog computing infrastructure in over 200 BS locations (including GSM and UMTS BSs) deployed by a primary MNO in the city of Dublin\cite{Francesco2018Dublin, Kibilda2016Dublin}. The locations and coverage areas of the BSs are presented in Figure \ref{Figure_BSdistribution}(a). To compare the workload offloading performance with different deployment densities of BSs, we consider 5 areas from the city center to the rural areas as shown in Figure \ref{Figure_BSdistribution}(b). We assume a mini-server consisting of 100 processing units that can be activated or deactivated according to the energy availability has been built inside of each BS. We assume each fog node can support two types of services: image and voice recognition with maximum tolerable response-time of 50 ms and 100 ms, respectively. \blu{These values are typical for today's voice and image recognition-based applications\cite{Assefi2015VoiceRecog, Kyllonen2016ResponsetimeImpact}.} Each processing unit can process 10 image or 40 voice recognition requests per second. \blu{We assume the number of energy units that can be harvested by each fog node follows a uniform distribution between 0 and a given value.} 
We consider two settings of offload forwarding. In the first setting, each fog node can only cooperate with its closest fog node. In the second one, each fog node can forward its workload to any neighboring fog nodes within a limited distance called (offload) forwarding distance. We assume the round-trip time between two fog nodes within the workload forwarding distance can be regarded as a constant given by $\tau_{ij} = 20ms$. We investigate the fog computing system under backlogged traffic conditions. 


To evaluate the performance improvement that can be achieved by offload forwarding, we compare the number of requests that can be processed by each BS in the five considered areas shown in Figure \ref{Figure_OffloadedWorkloadVSNumberofAreas}. We can observe that by allowing each BS to cooperate with all the neighboring BSs within a 500 meters of forwarding distance can significantly improve the numbers of offloaded requests for both supported services. We can also observe that even when each fog node can only cooperate with its closest fog node, the workload processing capability in terms of the number of offloaded requests can almost doubled compared to the case without offload forwarding. The workload processing capability can be further improved if each fog node can cooperate with more neighboring nodes. Note that in Figure \ref{Figure_OffloadedWorkloadVSNumberofAreas}, we observe that in areas 4 and 5, allowing each fog node to cooperate with other fog nodes within 500 meters cannot achieve the highest workload offloading performance. This is because in these two rural areas, some fog nodes cannot have any other fog nodes located in the 500 meters.

It can be observed that the round-trip workload transmission latency between neighboring fog nodes directly affect the performance of the offload forwarding. In Figure \ref{Figure_FigOffloadWorkloadVSRoundTripTime}, we compare the average number of offloaded service requests over all the considered areas when fog nodes within the workload forwarding distance have different round-trip time between each other. We can again observe that allowing each fog node to cooperate with all the neighboring nodes within the forwarding distance achieves the best offloading performance compared to other strategies even when the transmission latency between fog nodes becomes large. This is because in our proposed dynamic network slicing, each fog node can carefully schedule the energy consumed for activating the computational resources at each time slot. In this case, when the transmission latency between fog nodes is small, each fog node is more willing to spend energy in processing the workload for others or forwarding workload to others. On the other hand, when the round-trip time between fog nodes becomes large,
always forwarding workload to other neighboring fog nodes will introduce higher transmission latency. In this case, fog nodes will reserve more energy for its own use instead of helping others. 

In Figure \ref{Figure_FigOffloadWorkloadVSArrivalRate}, we fixed the workload arrival rate of the voice service for each fog node and compare the workload that can be processed by the fog nodes when the arrival rate for the image services changes. We can observe that both offload forwarding settings can significantly improve the offloading performance for the video service. This is because when each fog node receives more workload for the image processing service, it will be more likely to seek help from other fog nodes and jointly offload this highly loaded service with others. We can also observe that each fog node is always willing to offload more workload for the voice service than the image service. This is because the voice service require less computational resources for each activated processing unit.  

We consider the effect of energy harvesting process on the workload offloading performance of fog nodes in Figure \ref{Figure_FigOffloadVSHarvestEnergy} where we investigate the workload offloading performance of the fog layer when the maximum energy that can be harvested by each fog node is changed. Note that we assume the number of energy units that can be harvested by each fog node in each time slot is upper bounded by a maximum value. Therefore, the maximum amount of energy that can be harvested by each fog node also reflects the average energy that can be obtained by fog nodes as well as the energy available to activate the local computational resources. We observe that when the harvested energy for each fog node is limited, allowing two or more fog nodes to help each other can significantly improve the workload offloading performance of the fog layer. As the amount of harvested energy increases, more computational resources can be activated and the queuing delay of fog nodes will be further reduced. 

Finally, in Figure \ref{Figure_FigOffloadVSIterations}, we present the convergence performance of our proposed dynamic network slicing with the learning algorithm introduced in Section \ref{Section_Game}. 
We observe that our proposed approach can quickly converge to the optimal network slicing structure. 
We also compare the performance of myopic network slicing in which each fog node only tries to maximize the current workload offloading performance without considering the future energy harvesting and workload arrival processes. We can observe that allowing each fog node to carefully schedule its energy usage can significantly improve the workload offloading capability for the fog computing networks.

\vspace{-0.1in}
\section{Conclusion and Future Works}
\label{Section_Conclusion}

We have proposed the concept of dynamic network slicing for energy-harvesting fog computing networks consisting of a set of fog nodes that can offload and/or forward its received workload to CDCs using the energy harvested from natural environment. In this concept, each fog node can offload multiple types of service with guaranteed QoS  using the computational resources activated by the harvested energy. The limited resources and uncertainty of the energy harvesting process restrict the total amount of workload that can be processed by each individual fog node. 
To capture the fact that each fog node cannot know the global information but can make autonomous decisions using local information, we developed a stochastic overlapping coalition-formation game-based model to investigate the workload offloading problems.  We observe that the workload processing capacity can be significantly improved if each fog node can  learn from its previous interactions with others and sequentially refine the knowledge about the environmental state and private information of others. We have introduced a distributed optimization algorithm and proved that this algorithm achieves the optimal network slicing policy. Finally, we have considered a cellular network system that supports fog computing with renewable-energy-supply as a case study to evaluate the performance improvement that can be brought by our proposed framework. 

\blu{Our work in this paper also opens multiple future directions.
One direction of our research is to extend the proposed dynamic network slicing framework into new 5G core network architecture based on the so-called service based architecture (SBA). The design objective of this new architecture is to enable a more flexible deployment in which emerging services will be able to register themselves and connect to existing network components without introducing any new interface. Another potential direction is to extend the dynamic network slicing framework into fog computing supported by hybrid energy sources including both harvested energy as well as the power grid. In this system, the FSP will have to balance the energy supply from the free but uncertain renewable energy sources and that from the more reliable but costly power grid. }

\section*{Acknowledgment}
The authors would like to thank Professor Luiz A. DaSilva and Dr. Jacek Kibilda at CONNECT, Trinity College Dublin to provide the BS location data of Dublin city.

\appendices
\blu{
\section{Derivation of Response-time Equation (\ref{eq_ResponseTime1FogNodeOfflForward})}
\label{AppendixA}
Equation (\ref{eq_ResponseTime1FogNodeOfflForward}) follows directly from the derivations in \cite{XY2017FogCompInfocom, Keller2014QDelay}. We provide a brief description here for the completeness of this paper. Let us first consider a simple M/M/1 queuing delay at a single fog node $i$ serving type $k$ service workload by the on-board processors. Suppose the maximum processing capacity is $w_{i} p^{(k)}_{i,t}$ and the  amount of workload that needs to be processed by fog node $i$ is given by $\alpha^{(k)} _{i,t} \lambda^{(k)}_{i,t}$. According to \cite{Stallings2001NetworkQoS}, the response-time of type $k$ service at fog node $i$ is given by:
    \begin{eqnarray}
    \pi^{(k)}_{i,t} \left( \alpha^{(k)} _{i,t} \right) 
    = {1 \over  {w_{i} p^{(k)}_{i,t} 
    } - \alpha^{(k)} _{i,t} \lambda^{(k)}_{i,t}}. 
    \label{eq_ResponseTime1FogNodeNoCloud}
    \end{eqnarray}
}

\blu{
Suppose fog node $i$ can forward part of its received workload to a set ${\cal C}_i$ of its neighboring fog nodes. In this case, fog node $i$ needs to divide the total amount of workload into $|{\cal C}_i|$ partitions each of which will be sent to a fog node in ${\cal C}_i$. Let $\alpha^{(k)}_{ii, t}$ be the portion of workload received by fog node $i$ to be processed by its on-board processors. Let $\alpha^{(k)}_{im, t}$ be the portion of workload received by fog node $i$ that will be sent to another neighboring fog node $m$ for processing where $m \neq i$ and $m\in {\cal C}_i$. We have $\sum_{j \in {\cC}_{i} \cup \{i\}} \alpha^{(k)}_{jm,t} = 1$. In this case, the total amount of workload that needs to be processed by fog node $m$ is given by $\sum_{j \in {\cC}_{i} \cup \{i\}} \alpha^{(k)}_{jm,t} \lambda^{(k)}_{j,t}$. Suppose the maximum processing capability of fog node $m$ for processing type $k$ service is given by $w_m p^{(k)}_{m,t}$, we can write the response-time experienced by fog node $i$ forwarding $\alpha^{(k)}_{im,t}$ portion of the received workload to fog node $m$ as
\begin{eqnarray}
\pi^{(k)}_{im,t} = \alpha^{(k)}_{im,t} \left( \tau_{im} + {1 \over  w_m p^{(k)}_{m,t} - \sum_{j \in {\cC}_{i} \cup \{i\}} \alpha^{(k)}_{jm,t} \lambda^{(k)}_{j,t} } \right).
\end{eqnarray}
}

\blu{
Combining all portions of workload processed by fog node $i$ itself and the neighboring fog nodes in ${\cal C}_i$, we can obtain the response-time equation in (\ref{eq_ResponseTime1FogNodeOfflForward}). This concludes the proof.
}

\section{Proof of Theorem \ref{Theorem_Optimal}}
\label{Proof_Optimal}
To prove Theorem \ref{Theorem_Optimal}, we need to first prove that the resource slicing sub-game can be considered as a special overlapping-coalition-formation game  satisfying  the property of convexity, that is, a coalition of players can obtain more reward when it joins a larger coalition. Let ${\cal E} \left( {\cal F} \right)$ be the set of all feasible resource slicing agreements agreed by set $\cal F$ of fog nodes. We use $\langle \bc^{\cal C}, \balpha^{\cC} \rangle$ to denote a slicing agreement mutually agreed by a subset ${\cal C}\subseteq {\cal F}$ of fog nodes. 
More formally, a resource slicing game is {\em convex}\cite[Definition 13]{Chalkiadakis2010OverlapCoalitioanGame} if for each ${\cal C} \subseteq {\cal F}$ and ${\cal N} \subset {\cal O} \subseteq {\cal F} \backslash {\cal C}$, the following condition holds: for any ${\bc^{\cal N}, \balpha^{\cal N}} \in {\cal E} \left( {\cal N} \right)$, any $\langle \bc^{\cal O}, \balpha^{\cal O} \rangle \in {\cal E} \left( {\cal O} \right)$, and any $\langle \bc^{{\cal N}\cup{\cal C}}, \balpha^{{\cal N}\cup{\cal C}} \rangle \in {\cal E} \left( {\cal N}\cup{\cal C} \right)$ that satisfies $\bvarpi_{i} \left( \bc^{{\cal N}\cup{\cal C}}, \balpha^{{\cal N}\cup{\cal C}} \right) \ge \bvarpi_{i} \left( {\bc^{\cal N}, \balpha^{\cal N}} \right)$, $\forall i \in {\cal N}$, there exists an outcome $\langle \bc^{{\cal O}\cup{\cal C}}, \balpha^{{\cal O}\cup{\cal C}} \rangle \in {\cal E} \left( {\cal O}\cup{\cal C} \right)$ such that  $\bvarpi_{i} \left( \bc^{{\cal O}\cup{\cal C}}, \balpha^{{\cal O}\cup{\cal C}} \right) \ge \bvarpi_{i} \left( {\bc^{\cal O}, \balpha^{\cal O}} \right)$, $\forall i \in {\cal O}$ and $\bvarpi_{i} \left( \bc^{{\cal O}\cup{\cal C}}, \balpha^{{\cal O}\cup{\cal C}} \right) \ge \bvarpi_{i} \left( \bc^{{\cal N}\cup{\cal C}}, \balpha^{{\cal N}\cup{\cal C}} \right)$, $\forall i \in {\cal C}$.

Let us now prove that our resource slicing sub-game is convex.
It can be observed that in the resource slicing sub-game different fog nodes have different sets of neighboring fog nodes. The more fog nodes can cooperate with each other, the more resources can be utilized by all the member fog nodes. We can also observe that problem (\ref{eq_MaxOffloadMultipleFogNodes_SlicingSubgame}) is a linear function of $\balpha_{i,t}$. In addition, as mentioned in Section \ref{Section_Game}, each fog node will only cooperate with a subset of its neighboring fog nodes if it cannot obtain a higher reward by forming a coalition with other subsets of fog nodes. Let us write the solution of problem (\ref{eq_MaxOffloadMultipleFogNodes_SlicingSubgame}) as $\langle \bc^{{\cal C}*}_i, \balpha^{{\cal C}*}_i \rangle$ when the maximum set of fog nodes that can slice their resource to support all types of services is given by ${\cal C}*$. We can apply the standard convex optimization method to prove that the solution $\varpi_{i,t}\left( \langle \bc^{{\cal C}*}_i, \balpha^{{\cal C}*}_i \rangle \right)$ satisfies the following properties:
\begin{eqnarray}
\varpi_{i,t}\left( \langle \bc^{{{\cal O}\cup{\cal C}}*}_i, \balpha^{{{\cal O}\cup{\cal C}}*}_i \rangle \right) &\ge& \varpi_{i}\left( \langle \bc^{{{\cal O}}*}_i, \balpha^{{{\cal O}}*}_i \rangle \right),  \nonumber \\
\varpi_{i,t}\left( \langle \bc^{{{\cal O}\cup{\cal C}}*}_i, \balpha^{{{\cal O}\cup{\cal C}}*}_i \rangle \right) &\ge& \varpi_{i}\left( \langle \bc^{{{\cal N}\cup{\cal C}}*}_i, \balpha^{{{\cal N}\cup{\cal C}}*}_i \rangle \right), \nonumber \\
&&\;\;\;\;\;\; \forall {\cal N} \subset {\cal O} \subseteq {\cal F} \backslash {\cal C}.
\end{eqnarray}
We can therefore claim that the network slicing game is convex. 

We can now use the following theorem given in \cite{Chalkiadakis2010OverlapCoalitioanGame} to prove the non-emptiness of the core for our resource slicing sub-game.
\begin{theorem}\cite[Theorem 3]{Chalkiadakis2010OverlapCoalitioanGame}
If an overlapping coalition formation game is convex, and the worth $v$ is continuous, bounded, and monotone, and the maximum number of partial coalitions that each fog node can be involved in is finite, then the core of the game non-empty.
\end{theorem}

We have already proved the convexity of the resource slicing sub-game. Also, the worth defined in (\ref{eq_WorthofSlice}) satisfies all the above conditions. Therefore, we can claim that the core of the resource slicing sub-game is non-empty.

From the definition of the core and following the same line as \cite{Chalkiadakis2010OverlapCoalitioanGame}, we can prove that a network slicing agreement $\langle \bc, \balpha \rangle$ is in the core if and only if
\begin{eqnarray}
\sum_{i\in {\cal F}} \varpi_{i} \left( \bc, \balpha \right) \ge v^* \left(\bc^{\cal F}\right),
\end{eqnarray}
where $v^* \left(\bc^{\cal F}\right)$ is the supremum of $v \left(\bc^{\cal F}\right)$. In other words, any outcome in the core also maximizes the social welfare. 

\section{Proof of Theorem \ref{Theorem_MainAlgorithm}}
\label{Proof_Theorm_MainAlgorithm}

We briefly describe the proof of Theorem \ref{Theorem_MainAlgorithm} as follows. We can observe that the belief updating scheme for each fog node in (\ref{eq_PhyStateBeliefUpdate}) follows the Markov property. That is, the updated belief function of fog node $i$ is only related to its belief, outcome state and action in the previous time slot. We can also observe that (\ref{eq_LongTermPayoff}) is equivalent to the Bellman equation for single-agent POMDP. In other words, if each fog node regards the environment as well as the decision making processes of the other fog nodes as part of the system state, the workload offloading problem for each fog node can be regarded as a single-agent POMDP. We can then follow the same line as \cite{Puterman2014MDP} to prove that (\ref{eq_OptimalPolicy}) is the optimal policy for each fog node to maximize its long-term workload offloading performance. 
We omit the details of the proof due to the limit of space. This concludes the proof.  %



\vspace{-0.2in}
\bibliography{reference}

\begin{thebibliography}{10}
\providecommand{\url}[1]{#1}
\csname url@samestyle\endcsname
\providecommand{\newblock}{\relax}
\providecommand{\bibinfo}[2]{#2}
\providecommand{\BIBentrySTDinterwordspacing}{\spaceskip=0pt\relax}
\providecommand{\BIBentryALTinterwordstretchfactor}{4}
\providecommand{\BIBentryALTinterwordspacing}{\spaceskip=\fontdimen2\font plus
\BIBentryALTinterwordstretchfactor\fontdimen3\font minus
  \fontdimen4\font\relax}
\providecommand{\BIBforeignlanguage}[2]{{%
\expandafter\ifx\csname l@#1\endcsname\relax
\typeout{** WARNING: IEEEtran.bst: No hyphenation pattern has been}%
\typeout{** loaded for the language `#1'. Using the pattern for}%
\typeout{** the default language instead.}%
\else
\language=\csname l@#1\endcsname
\fi
#2}}
\providecommand{\BIBdecl}{\relax}
\BIBdecl

\bibitem{ITU2014TactileInternet}
\BIBentryALTinterwordspacing
{ITU-T}, ``The {T}actile {I}nternet,'' {ITU}-{T} technology watch report, Aug.
  2014. [Online]. Available:
  \url{https://www.itu.int/dms_pub/itu-t/opb/gen/T-GEN-TWATCH-2014-1-PDF-E.pdf}
\BIBentrySTDinterwordspacing

\bibitem{Huawei5GVision}
\BIBentryALTinterwordspacing
Huawei, ``5{G} vision: 100 billion connections, 1 ms latency, and 10 gbps
  throughput.'' [Online]. Available:
  \url{https://www.huawei.com/minisite/5g/en/defining-5g.html}
\BIBentrySTDinterwordspacing

\bibitem{Andrews5G}
J.~G. Andrews, S.~Buzzi, W.~Choi, S.~V. Hanly, A.~Lozano, A.~C.~K. Soong, and
  J.~C. Zhang, ``What will 5{G} be?'' \emph{IEEE Journal on Selected Areas in
  Communications}, vol.~32, no.~6, pp. 1065--1082, Jun. 2014.

\bibitem{Vaquero2014FogComp}
L.~Vaquero and L.~Rodero-Merino, ``Finding your way in the fog: Towards a
  comprehensive definition of fog computing,'' \emph{Proc. of ACM SIGCOMM
  Comput. Commun. Rev.}, vol.~44, no.~5, pp. 27--32, Oct. 2014.

\bibitem{NGMN5GWhitePaper}
\BIBentryALTinterwordspacing
{NGMN Alliance}, ``5{G} white paper,'' Feb. 2015. [Online]. Available:
  \url{https://www.ngmn.org/uploads/media/NGMN_5G_White_Paper_V1_0.pdf}
\BIBentrySTDinterwordspacing

\bibitem{Dastjerdi2016FogComp}
A.~{Vahid Dastjerdi}, H.~{Gupta}, R.~N. {Calheiros}, S.~K. {Ghosh}, and
  R.~{Buyya}, ``{Fog Computing: Principals, Architectures, and Applications},''
  \emph{ArXiv e-prints}, Jan. 2016.

\bibitem{Yi2015Fog}
S.~Yi, C.~Li, and Q.~Li, ``A survey of fog computing: Concepts, applications
  and issues,'' in \emph{Proc. of ACM Workshop on Mobile Big Data}, Hangzhou,
  China, Jun. 2015, pp. 37--42.

\bibitem{Yannuzzi2014Fog}
M.~Yannuzzi, R.~Milito, R.~Serral-Gracia, D.~Montero, and M.~Nemirovsky, ``Key
  ingredients in an iot recipe: Fog computing, cloud computing, and more fog
  computing,'' in \emph{Proc. of the IEEE International Workshop on Computer
  Aided Modeling and Design of Communication Links and Networks}, Athens, Dec.
  2014, pp. 325--329.

\bibitem{GoogleGreenCloud}
\BIBentryALTinterwordspacing
``Google cloud and the environment,'' Google. [Online]. Available:
  \url{https://cloud.google.com/environment/}
\BIBentrySTDinterwordspacing

\bibitem{AppleGreenCloud}
\BIBentryALTinterwordspacing
``Apple becomes a green energy supplier, with itself as customer,'' New York
  Times, Aug. 2016. [Online]. Available:
  \url{https://www.nytimes.com/2016/08/24/business/energy-environment/as-energy-use-rises-corporations-turn-to-their-own-green-utility-sources.html}
\BIBentrySTDinterwordspacing

\bibitem{MicrosoftGreenCloud}
\BIBentryALTinterwordspacing
``Microsoft environment: Enabling a sustainable future,'' Microsoft. [Online].
  Available: \url{https://www.microsoft.com/en-us/environment/default.aspx}
\BIBentrySTDinterwordspacing

\bibitem{FacebookGreenCloud}
\BIBentryALTinterwordspacing
``Apple, facebook, and google top greenpeace energy report card,'' Fortune.com.
  [Online]. Available:
  \url{http://fortune.com/2017/01/10/greenpeace-energy-report-apple-facebook-google/}
\BIBentrySTDinterwordspacing

\bibitem{Chamola2016SolarBS}
V.~Chamola and B.~Sikdar, ``Solar powered cellular base stations: current
  scenario, issues and proposed solutions,'' \emph{IEEE Communications
  Magazine}, vol.~54, no.~5, pp. 108--114, May 2016.

\bibitem{Ulukus2015EHreview}
S.~Ulukus, A.~Yener, E.~Erkip, O.~Simeone, M.~Zorzi, P.~Grover, and K.~Huang,
  ``Energy harvesting wireless communications: A review of recent advances,''
  \emph{IEEE J. Sel. Areas in Commun.}, vol.~33, no.~3, pp. 360--381, Mar.
  2015.

\bibitem{XY2015DET}
Y.~Xiao, D.~Niyato, Z.~Han, and L.~DaSilva, ``Dynamic energy trading for energy
  harvesting communication networks: A stochastic energy trading game,''
  \emph{IEEE J. Sel. Areas Commun.}, vol.~33, no.~12, pp. 2718--2734, Dec.
  2015.

\bibitem{Lu2015EHSurvey}
X.~Lu, P.~Wang, D.~Niyato, D.~I. Kim, and Z.~Han, ``Wireless networks with {RF}
  energy harvesting: A contemporary survey,'' \emph{IEEE Communications Surveys
  Tutorials}, vol.~17, no.~2, pp. 757--789, 2015.

\bibitem{XY2015ICCBayeRL}
Y.~Xiao, Z.~Han, D.~Niyato, and C.~Yuen, ``Bayesian reinforcement learning for
  energy harvesting communication systems with uncertainty,'' in \emph{Proc. of
  the IEEE ICC Conference}, London, UK, Jun. 2015.

\bibitem{Ge2015EnergyEff}
X.~Ge, B.~Yang, J.~Ye, G.~Mao, C.~Wang, and T.~Han, ``Spatial spectrum and
  energy efficiency of random cellular networks,'' \emph{IEEE Trans. Commun.},
  vol.~63, no.~3, pp. 1019--1030, Mar 2015.

\bibitem{3GPP2016NetworkShare}
3GPP, ``Telecommunication management; network sharing; concepts and
  requirements,'' 3GPP TS 32.130, Jun. 2016.

\bibitem{3GPP2016NetworkShare2}
------, ``Network sharing; artechecture and functional description,'' 3GPP TR
  23.251, Jun. 2016.

\bibitem{Vaezi2017VirtualizationCloud}
\BIBentryALTinterwordspacing
M.~Vaezi and Y.~Zhang, \emph{Virtualization and Cloud Computing}.\hskip 1em
  plus 0.5em minus 0.4em\relax Springer, 2017, pp. 11--31. [Online]. Available:
  \url{https://doi.org/10.1007/978-3-319-54496-0_2}
\BIBentrySTDinterwordspacing

\bibitem{NGMN2016NetworkSlicing}
\BIBentryALTinterwordspacing
{NGMN Alliance}, ``Description of network slicing concept,'' Sep. 2016.
  [Online]. Available:
  \url{https://www.ngmn.org/uploads/media/161010_NGMN_Network_Slicing_framework_v1.0.8.pdf}
\BIBentrySTDinterwordspacing

\bibitem{Richart2016NetSlicing}
M.~Richart, J.~Baliosian, J.~Serrat, and J.~L. Gorricho, ``Resource slicing in
  virtual wireless networks: A survey,'' \emph{IEEE Transactions on Network and
  Service Management}, vol.~13, no.~3, pp. 462--476, Sep. 2016.

\bibitem{Samdanis2016NetworkSlicingBroker}
K.~Samdanis, X.~Costa-Perez, and V.~Sciancalepore, ``From network sharing to
  multi-tenancy: The 5g network slice broker,'' \emph{IEEE Communications
  Magazine}, vol.~54, no.~7, pp. 32--39, July 2016.

\bibitem{ONF5GSlicing}
ONF, ``Applying {SDN} architecture to 5{G} slicing issue 1,'' ONF TR-526, Apr.
  2016.

\bibitem{ONFSDNArchitecture}
------, ``{SDN} architecture issue 1.1,'' ONF TR-521, 2016.

\bibitem{Sarkar2016Fog}
\BIBentryALTinterwordspacing
S.~Sarkar, S.~Chatterjee, and S.~Misra, ``Assessment of the suitability of fog
  computing in the context of internet of things,'' \emph{to appear at IEEE
  Transactions on Cloud Computing}. [Online]. Available:
  \url{http://ieeexplore.ieee.org/document/7286781/}
\BIBentrySTDinterwordspacing

\bibitem{Bonomi2014Fog}
F.~Bonomi, R.~Milito, P.~Natarajan, and J.~Zhu, \emph{Big Data and Internet of
  Things: A Roadmap for Smart Environments}.\hskip 1em plus 0.5em minus
  0.4em\relax Springer, 2014, ch. Fog Computing: A Platform for Internet of
  Things and Analytics, pp. 169--186.

\bibitem{Datta2015Fog}
S.~Datta, C.~Bonnet, and J.~Haerri, ``Fog computing architecture to enable
  consumer centric internet of things services,'' in \emph{Proc. of IEEE ISCE
  Conference}, Madrid, Spain, Jun. 2015, pp. 1--2.

\bibitem{Do2015Fog}
C.~Do, N.~Tran, C.~Pham, M.~Alam, J.~H. Son, and C.~S. Hong, ``A proximal
  algorithm for joint resource allocation and minimizing carbon footprint in
  geo-distributed fog computing,'' in \emph{Proc. of the IEEE ICOIN
  Conference}, Jan Siem Reap, Cambodia, Jan. 2015, pp. 324--329.

\bibitem{Aazam2015Fog}
M.~Aazam and E.-N. Huh, ``Dynamic resource provisioning through fog micro
  datacenter,'' in \emph{Proc. of the IEEE PerCom Workshops}, St. Louis, MO,
  Mar. 2015, pp. 105--110.

\bibitem{XY2018TactielInternet}
Y.~Xiao and M.~Krunz, ``Distributed optimization for energy-efficient fog
  computing in the tactile internet,'' \emph{to be published in IEEE J. Sel.
  Areas Commun.}, 2018.

\bibitem{Harshit2016SDFog}
\BIBentryALTinterwordspacing
H.~Gupta, S.~B. Nath, S.~Chakraborty, and S.~K. Ghosh, ``Sdfog: {A} software
  defined computing architecture for qos aware service orchestration over edge
  devices,'' \emph{CoRR}, vol. abs/1609.01190, 2016. [Online]. Available:
  \url{http://arxiv.org/abs/1609.01190}
\BIBentrySTDinterwordspacing

\bibitem{Khan2018VANETFog}
A.~A. Khan, M.~Abolhasan, and W.~Ni, ``5g next generation vanets using sdn and
  fog computing framework,'' in \emph{Proc of IEEE CCNC Conference}, Las Vegas,
  NV, Jan. 2018, pp. 1--6.

\bibitem{Hakiri2017Fog}
A.~Hakiri, B.~Sellami, P.~Patil, P.~Berthou, and A.~Gokhale, ``Managing
  wireless fog networks using software-defined networking,'' in \emph{Proc. of
  IEEE/ACS AICCSA Conference}, Hammamet, Tunisia, Oct. 2017, pp. 1149--1156.

\bibitem{Tootoonchian2010SDNHyperflow}
A.~Tootoonchian and Y.~Ganjali, ``Hyperflow: A distributed control plane for
  {O}pen{F}low,'' in \emph{Proceedings of the 2010 Internet Network Management
  Workshop/Workshop on Research on enterprise networking}, San Jose, CA, Apr.
  2010, pp. 1--6.

\bibitem{Fang2015HierarchicalSDN}
\BIBentryALTinterwordspacing
L.~Fang, F.~Chiussi, D.~Bansal, V.~Gill, T.~Lin, J.~Cox, and G.~Ratterree,
  ``Hierarchical sdn for the hyper-scale, hyper-elastic data center and
  cloud,'' in \emph{Proceedings of the 1st ACM SIGCOMM Symposium on Software
  Defined Networking Research}.\hskip 1em plus 0.5em minus 0.4em\relax New
  York, NY, USA: ACM, Jun. 2015, pp. 7:1--7:13. [Online]. Available:
  \url{http://doi.acm.org/10.1145/2774993.2775009}
\BIBentrySTDinterwordspacing

\bibitem{Yeganeh2012KandooSDN}
\BIBentryALTinterwordspacing
S.~Hassas~Yeganeh and Y.~Ganjali, ``Kandoo: A framework for efficient and
  scalable offloading of control applications,'' in \emph{Proceedings of the
  1st Workshop on Hot Topics in Software Defined Networks}, Aug. 2012, pp.
  19--24. [Online]. Available: \url{http://doi.acm.org/10.1145/2342441.2342446}
\BIBentrySTDinterwordspacing

\bibitem{Liu2015HierarchicalSDN}
Y.~Liu, A.~Hecker, R.~Guerzoni, Z.~Despotovic, and S.~Beker, ``On optimal
  hierarchical sdn,'' in \emph{Proc of IEEE ICC Conference}, June London, UK,
  Jun. 2015, pp. 5374--5379.

\bibitem{ZhangHQ2016FogComp}
H.~Zhang, Y.~Xiao, S.~Bu, D.~Niyato, R.~Yu, and Z.~Han, ``Fog computing in
  multi-tier data center networks: A hierarchical game approach,'' in
  \emph{Proc. of the IEEE ICC Conference}, Kuala Lumpur, Malaysia, May 2016.

\bibitem{Chiang2017FogBook}
M.~Chiang, B.~Balasubramanian, and F.~Bonomi, \emph{Fog for 5G and IoT}.\hskip
  1em plus 0.5em minus 0.4em\relax John Wiley \& Sons, 2017.

\bibitem{Zhang2017FogCompIoT}
H.~Zhang, Y.~Xiao, S.~Bu, D.~Niyato, F.~R. Yu, and Z.~Han, ``Computing resource
  allocation in three-tier iot fog networks: a joint optimization approach
  combining stackelberg game and matching,'' \emph{IEEE Internet of Things
  Journal}, vol.~4, no.~5, pp. 1204--1215, 2017.

\bibitem{Zhang2017FogCompTCC}
H.~Zhang, Y.~Xiao, S.~Bu, R.~Yu, D.~Niyato, and Z.~Han, ``Distributed resource
  allocation for data center networks: A hierarchical game approach,'' \emph{to
  appear at IEEE Transactions on Cloud Computing}.

\bibitem{Tong2016}
L.~Tong, Y.~Li, and W.~Gao, ``A hierarchical edge cloud architecture for mobile
  computing,'' in \emph{Proc. of IEEE INFOCOM Conf.}, San Francisco, CA, Apr.
  2016.

\bibitem{Hadzic2017EdgeComp}
\BIBentryALTinterwordspacing
I.~Had\v{z}i\'{c}, Y.~Abe, and H.~C. Woithe, ``Edge computing in the epc: A
  reality check,'' in \emph{Proc. of the ACM/IEEE Symposium on Edge
  Computing}.\hskip 1em plus 0.5em minus 0.4em\relax New York, NY, USA: ACM,
  2017, pp. 13:1--13:10. [Online]. Available:
  \url{http://doi.acm.org/10.1145/3132211.3134449}
\BIBentrySTDinterwordspacing

\bibitem{Sciancalepore2017NetSlice}
V.~Sciancalepore, K.~Samdanis, and X.~Costa-Perez, ``Mobile traffic forecasting
  for maximizing 5g network slicing resource utilization,'' in \emph{Proc. of
  the IEEE INFOCOM Conference}, Atlanta, GA, May 2017.

\bibitem{Caballero2017NetSlicingGame}
P.~Caballero, A.~Banchs, G.~de~Veciana, and X.~Costa-Perez, ``Network slicing
  games: Enabling customization in multi-tenant networks,'' in \emph{Proc. of
  the IEEE INFOCOM Conference}, Atlanta, GA, May 2017.

\bibitem{Bega2017NetworkSlicing}
D.~Bega, M.~Gramaglia, A.~Banchs, V.~Sciancalepore, K.~Samdanis, and
  X.~Costa-Perez, ``Optimising 5g infrastructure markets: The business of
  network slicing,'' in \emph{Proc. of the IEEE INFOCOM Conference}, Atlanta,
  GA, May 2017.

\bibitem{ATT2017EdgeNetworks}
\BIBentryALTinterwordspacing
``The cloud comes to you: {AT\&T} to power self-driving cars, {AR/VR} and other
  future 5{G} applications through edge computing,'' AT\&T Newsroom, Jul. 2017.
  [Online]. Available:
  \url{http://about.att.com/story/reinventing_the_cloud_through_edge_computing.html}
\BIBentrySTDinterwordspacing

\bibitem{IETF2017ARVRResponsetime}
\BIBentryALTinterwordspacing
``Problem statement: Transport support for augmented and virtual reality
  applications,'' Internet-Draft, {IETF}, Mar. 2017. [Online]. Available:
  \url{https://tools.ietf.org/id/draft-han-iccrg-arvr-transport-problem-00.xml}
\BIBentrySTDinterwordspacing

\bibitem{Aprem2013EnergHarv}
A.~Aprem, C.~Murthy, and N.~Mehta, ``Transmit power control policies for energy
  harvesting sensors with retransmissions,'' \emph{IEEE J. Sel. Topics in
  Signal Process.}, vol.~7, no.~5, pp. 895--906, Oct. 2013.

\bibitem{Keller2014QDelay}
\BIBentryALTinterwordspacing
M.~Keller and H.~Karl, ``Response time-optimized distributed cloud resource
  allocation,'' \emph{arXiv preprint arXiv:1601.06262}, 2016. [Online].
  Available: \url{http://arxiv.org/abs/1601.06262}
\BIBentrySTDinterwordspacing

\bibitem{Deng2016FogCompu}
R.~Deng, R.~Lu, C.~Lai, T.~H. Luan, and H.~Liang, ``Optimal workload allocation
  in fog-cloud computing toward balanced delay and power consumption,''
  \emph{IEEE Internet of Things Journal}, vol.~3, no.~6, pp. 1171--1181, Dec
  2016.

\bibitem{XY2014GC14CoopEnergHarvest}
Y.~Xiao, Z.~Han, and L.~A. DaSilva, ``Opportunistic relay selection for
  cooperative energy harvesting communication networks,'' in \emph{Proc. of the
  IEEE GLOBECOM}, Austin, TX, Dec. 2014.

\bibitem{Chia2014RenewableBSs}
Y.~K. Chia, S.~Sun, and R.~Zhang, ``Energy cooperation in cellular networks
  with renewable powered base stations,'' \emph{IEEE Transactions on Wireless
  Communications}, vol.~13, no.~12, pp. 6996--7010, Dec 2014.

\bibitem{Li2016RenewableBSs}
Q.~Li, Y.~Wei, M.~Song, and F.~R. Yu, ``Traffic aware energy management in
  cellular networks with renewable energy powered base stations,'' in
  \emph{Proc. of the IEEE VTC Spring Conference}, May 2016, pp. 1--5.

\bibitem{Zhang2015DataCenter}
L.~Zhang, S.~Ren, C.~Wu, and Z.~Li, ``A truthful incentive mechanism for
  emergency demand response in colocation data centers,'' in \emph{Proc. of the
  IEEE INFOCOM Conference}, Hong Kong, China, Apr. 2015, pp. 2632--2640.

\bibitem{Tran2015DataCenter}
N.~Tran, C.~Do, S.~Ren, Z.~Han, and C.~S. Hong, ``Incentive mechanisms for
  economic and emergency demand responses of colocation datacenters,''
  \emph{IEEE J. Sel. Areas in Commun.}, vol.~33, no.~12, pp. 2892--2905, Dec.
  2015.

\bibitem{XY2017FogCompInfocom}
Y.~Xiao and M.~Krunz, ``Qo{E} and power efficiency tradeoff for fog computing
  networks with fog node cooperation,'' in \emph{Proc. of the IEEE INFOCOM
  Conference}, Atlanta, GA, May 2017.

\bibitem{Chalkiadakis2010OverlapCoalitioanGame}
G.~Chalkiadakis, E.~Elkind, E.~Markakis, M.~Polukarov, and N.~R. Jennings,
  ``Cooperative games with overlapping coalitions,'' \emph{Journal of
  Artificial Intelligence Research}, vol.~39, no.~1, pp. 179--216, Sep. 2010.

\bibitem{XY2015GlobecomEnergHarvest}
Y.~Xiao, D.~Niyato, Z.~Han, and L.~A. DaSilva, ``Joint optimization for power
  scheduling and transfer in energy harvesting communication systems,'' in
  \emph{Proc. of the IEEE GLOBECOM Conference}, San Diego, CA, Dec. 2015.

\bibitem{Chalkiadakis2010SequentialCoalitioanGame}
G.~Chalkiadakis and C.~Boutilier, ``Sequentially optimal repeated coalition
  formation under uncertainty,'' \emph{Autonomous Agents and Multi-Agent
  Systems}, vol.~24, no.~3, pp. 441--484, May 2012.

\bibitem{Gmytrasiewicz2005IPOMDPs}
P.~Gmytrasiewicz and P.~Doshi, ``A framework for sequential planning in
  multiagent settings,'' \emph{Journal of Artificial Intelligence Research},
  vol.~24, no.~1, pp. 49--79, Jul. 2005.

\bibitem{Francesco2018Dublin}
P.~D. Francesco, F.~Malandrino, and L.~A. DaSilva, ``Assembling and using a
  cellular dataset for mobile network analysis and planning,'' \emph{to appear
  at IEEE Transactions on Big Data}, 2018.

\bibitem{Kibilda2016Dublin}
J.~Kibilda, B.~Galkin, and L.~A. DaSilva, ``Modelling multi-operator base
  station deployment patterns in cellular networks,'' \emph{IEEE Transactions
  on Mobile Computing}, vol.~15, no.~12, pp. 3087--3099, Dec 2016.

\bibitem{Assefi2015VoiceRecog}
M.~Assefi, M.~Wittie, and A.~Knight, ``Impact of network performance on cloud
  speech recognition,'' in \emph{Proc of the ICCCN Conference}, Las Vegas, NV,
  Aug. 2015, pp. 1--6.

\bibitem{Kyllonen2016ResponsetimeImpact}
P.~C. Kyllonen and J.~Zu, ``Use of response time for measuring cognitive
  ability,'' \emph{Journal of Intelligence}, vol.~4, no.~4, p.~14, Apr. 2016.

\bibitem{Stallings2001NetworkQoS}
W.~Stallings, \emph{High Speed Networks and Internets: Performance and Quality
  of Service}, 2nd~ed.\hskip 1em plus 0.5em minus 0.4em\relax Upper Saddle
  River, NJ, USA: Prentice Hall PTR, 2001.

\bibitem{Puterman2014MDP}
M.~Puterman, \emph{Markov Decision Processes: Discrete Stochastic Dynamic
  Programming}, ser. Wiley Series in Probability and Statistics, 2005.

\end{thebibliography}
\bibliographystyle{IEEEtran}

\begin{IEEEbiography}{Yong Xiao}(S'09-M'13-SM'15) received his B.S. degree in electrical engineering from China University of Geosciences, Wuhan, China in 2002, M.Sc. degree in telecommunication from Hong Kong University of Science and Technology in 2006, and his Ph. D degree in electrical and electronic engineering from Nanyang Technological University, Singapore in 2012.
From August 2010 to April 2011, he was a Research Associate in School of Electrical and Electronic Engineering, Nanyang Technological University, Singapore. From May 2011 to
October 2012, he was a Research Fellow at CTVR, School of Computer Science and Statistics, Trinity College Dublin, Ireland. From November 2012 to December 2013, he was a Postdoctoral Fellow at Massachusetts Institute of Technology. From December 2013 to November 2014, he was an MIT-SUTD
Postdoctoral Fellow with Singapore University of Technology and Design and Massachusetts Institute of Technology. From November 2014 to August 2016, he was a Postdoctoral Fellow II with the Department of Electrical and Computer Engineering, University of Houston. From September 2016 to September 2018, he was a  research assistant professor in the Department of Electrical and Computer Engineering at the University of Arizona where he was also the center manager
of the Broadband Wireless Access and Applications Center (BWAC), an NSF Industry/University Cooperative
Research Center (I/UCRC) led by the University of Arizona.
Currently, he is a professor in the School of Electronic Information and Communications at the Huazhong University of Science and Technology (HUST), Wuhan, China. His research interests include machine learning, game theory, distributed optimization, and their applications in cloud/fog/mobile edge computing, green communication systems, wireless communication networks, and Internet-of-Things (IoT).
\end{IEEEbiography}


\vskip 0pt plus -1fil

\begin{IEEEbiography}{Marwan Krunz}(S'93-M'95-SM'04-F'10) is the Kenneth VonBehren Endowed Professor in the Department of Electrical and Computer Engineering at the University of Arizona (UA). He also holds a joint (courtesy) appointment as a Professor in the Department of Computer Science. From 2008 to 2013, he was the UA site director of ``Connection One", an NSF Industry/University Cooperative Research Center (I/UCRC) that focused on RF and wireless communication systems and networks. During that period, the center included five participating sites (ASU, UA, OSU, RPI, and the University of Hawaii) and 26+ affiliates from industry and national research labs. Currently, Dr. Krunz co-directs the Broadband Wireless Access and Applications Center (BWAC), an NSF I/UCRC that includes UA (lead site), Virginia Tech, University of Notre Dame, University of Mississippi, Auburn University, and Catholic University of America. Along with NSF support, BWAC is currently funded by numerous companies and DoD labs, including Raytheon, Keysight Technologies, Alcatel-Lucent, Intel, L-3 Communications, Motorola Solutions, National Instruments, ONR, and others. BWAC aims at advancing the underlying technologies and providing cost-effective and practical solutions for next-generation (5G \& beyond) wireless systems, millimeter-wave communications, wireless cybersecurity, shared-spectrum access systems, full-duplex transmissions, massive MIMO techniques, and others. Dr. Krunz received the Ph.D. degree in electrical engineering from Michigan State University in July 1995. He joined the University of Arizona in January 1997, after a brief postdoctoral stint at the University of Maryland, College Park. In 2010, he was a Visiting Chair of Excellence (``Catedra de Excelencia") at the University of Carlos III de Madrid (Spain), and concurrently a visiting researcher at Institute IMDEA Networks. In summer 2011, he was a Fulbright Senior Specialist, visiting with the University of Jordan, King Abdullah II School of Information Technology. He previously held numerous other short-term research positions at the University Technology Sydney, Australia (2016), University of Paris V (2013), INRIA-Sophia Antipolis, France (2011, 2008, and 2003), University of Paris VI (LIP6 Group, 2006), HP Labs, Palo Alto (2003), and US West Advanced Technologies (1997).
\end{IEEEbiography}


\end{document}